\documentclass[twocolumn]{aastex6}
\usepackage[flushleft]{threeparttable}
\usepackage{ccaption}

\fullcollaborationName{ Co-authors and affiliations can be found after the acknowledgements}

\newcommand{\GAIA}{{\it Gaia }}

\begin{document}


\shorttitle{A chemodynamical study of the Open Cluster IC 166}
\shortauthors{J. Schiappacasse-Ulloa et al.}

\title{A chemical and kinematical analysis of the intermediate-age open cluster IC 166 from APOGEE and \GAIA DR2}


\author{
	   J. Schiappacasse-Ulloa\altaffilmark{1},
	  B. Tang\altaffilmark{2, 1},
	  J. G. Fern\'andez-Trincado\altaffilmark{1, 3}, 
	  O. Zamora\altaffilmark{4, 5},	  
	  D. Geisler\altaffilmark{1},
	  P. Frinchaboy\altaffilmark{6},
	  M. Schultheis\altaffilmark{7}, 
	  F. Dell'Agli\altaffilmark{4, 5},  
	  S. Villanova\altaffilmark{1},
	  T. Masseron\altaffilmark{4, 5},
	  Sz. M\'esz\'aros\altaffilmark{9, 10},
	  D. Souto\altaffilmark{11},  	  
	  S. Hasselquist\altaffilmark{12},	  
	  K. Cunha\altaffilmark{11,13},  
	  V. V. Smith\altaffilmark{20}, 
	  D. A. Garc\'ia-Hern\'andez\altaffilmark{4, 5},
	  K. Vieira\altaffilmark{21},   	   
	  A. C. Robin\altaffilmark{3},  
	  D. Minniti\altaffilmark{14, 15, 16},
	  G. Zasowski\altaffilmark{19},	
	  E. Moreno\altaffilmark{17}, 
	  A. P\'erez-Villegas\altaffilmark{18},	 
	  R. R. Lane\altaffilmark{8},
	  I. I. Ivans\altaffilmark{19},
	  K. Pan\altaffilmark{23},
	  C. Nitschelm\altaffilmark{24},
	  F. A. Santana\altaffilmark{25},
	  R. Carrera\altaffilmark{4,5},
	  \&
	  A. Roman-Lopes\altaffilmark{26}
	  }
	  
	  \affil{
	  $^{1}$Departamento de Astronom\'ia, Univerisidad de Concepci\'on, Av. Esteban Iturra s/n Barrio Universitario, Casilla 160-C Concepci\'on, Chile: \href{mailto:}{jfernandezt@astro-udec.cl, jfernandezt87@gmail.com, jschiappacasseu@gmail.com}\\
	  $^{2}$School of Physics and Astronomy, Sun Yat-sen University, Zhuhai 519082, China: \href{mailto:}{tangbt@mail.sysu.edu.cn}\\
	  $^{3}$Institut Utinam, CNRS UMR6213, Univ. Bourgogne Franche-Comt\'e, OSU THETA , Observatoire de Besan\c{}con, BP 1615, 25010 Besan\c{}con Cedex, France: \href{mailto:}{jfernandez@obs-besancon.fr}\\
	  $^{4}$Instituto de Astrof\'isica de Canarias, V\'ia L\'actea, 38205 La Laguna, Tenerife, Spain\\
	  $^{5}$Universidad de La Laguna, Departamento de Astrof\'isica, 38206 La Laguna, Tenerife, Spain\\
	  $^{6}$Department of Physics and Astronomy, Texas Christian University, Fort Worth, TX 76129, USA\\
	  $^{7}$Laboratoire Lagrange, Universit\'e C\^{o}te d'Azur, Observatoire de la C\^{o}te d'Azur, CNRS, Bd de l'Observatoire, 06304 Nice, France\\
	  $^{8}$Instituto de Astrof\'isica, Pontificia Universidad Cat\'olica de Chile, Av. Vicu\~na Mackenna 4860, 782-0436 Macul, Santiago, Chile\\
	  $^{9}$ELTE E\"{o}tv\"{o}s Lor\'and University, Gothard Astrophysical Observatory, Szombathely, Hungary\\
	  $^{10}$Premium Postdoctoral Fellow of the Hungarian Academy of Sciences\\
	  $^{11}$Observat\'orio Nacional, 20921-400 Sao Crist\'ovao, Rio de Janeiro, Brazil\\
	  $^{12}$New Mexico State University, Las Cruces, NM 88003, USA\\
	  $^{13}$Steward Observatory, University of Arizona, 933 North Cherry Avenue, Tucson, AZ 85721, USA\\
	  $^{14}$Departamento de Fisica, Facultad de Ciencias Exactas, Universidad Andres Bello Av. Fernandez Concha 700,
	  7591538 Las Condes, Santiago, Chile\\
	  $^{15}$Instituto Milenio de Astrof\'isica, Santiago, Chile\\
	  $^{16}$Vatican Observatory, V00120 Vatican City State, Italy\\
	  $^{17}$Instituto de Astronom\'ia, Universidad Nacional Aut\'onoma de M\'exico, Apdo. Postal 70264, M\'exico D.F., 04510, M\'exico\\
	  $^{18}$Universidade de S\~ao Paulo, IAG, Rua do Mat\~ao 1226, Cidade Universit\'aria, 05508-900, S\~ao Paulo, Brazil\\
	  $^{19}$Department of Physics and Astronomy, The University of Utah, Salt Lake City, UT 84112, USA\\
	  $^{20}$National Optical Astronomy Observatories, Tucson, AZ 85719, USA\\
	  $^{21}$Centro de Investigaciones de Astronom\'ia, AP 264,M\'erida 5101-A, Venezuela\\
	  $^{22}$Department of Astronomy, University of Virginia, Charlottesville, VA 22904-4325, USA\\
	  $^{23}$Apache Point Observatory and New Mexico State University, P.O. Box 59, Sunspot, NM, 88349-0059, USA\\
	  $^{24}$Unidad de Astronom\'ia, Universidad de Antofagasta, Avenida Angamos 601, Antofagasta 1270300, Chile\\
	  $^{25}$Universidad de Chile, Av. Libertador Bernardo O'Higgins 1058, Santiago de Chile\\
	  $^{26}$Departamento de F\'isica, Facultad de Ciencias, Universidad de La Serena, Cisternas 1200, La Serena, Chile
	  }




\begin{abstract}
IC 166 is an intermediate-age open cluster ($\sim 1$ Gyr) which lies in the transition zone of the metallicity gradient in the outer disc. Its location, combined with our very limited knowledge of its salient features, make it an interesting object of study. We present the first high-resolution spectroscopic and precise kinematical analysis of IC 166, which lies in the outer disc with  $R_{GC} \sim 12.7$ kpc. High resolution \textit{H}-band spectra were analyzed using observations from the SDSS-IV Apache Point Observatory Galactic Evolution Experiment (APOGEE) survey. We made use of the Brussels Automatic Stellar Parameter (BACCHUS) code to provide chemical abundances based on a line-by-line approach for up to eight chemical elements (Mg, Si, Ca, Ti, Al, K, Mn and Fe). The $\alpha-$element (Mg, Si, Ca and whenever available Ti) abundances, and their trends with Fe abundances have been analysed for a total of 13 high-likelihood cluster members.  No significant abundance scatter was found in any of the chemical species studied. Combining the positional, heliocentric distance, and kinematic information we derive, for the first time, the probable orbit of IC 166 within a Galactic model including a rotating boxy bar, and found that it is likely that IC 166 formed in the Galactic disc, supporting its nature as an unremarkable Galactic open cluster with an orbit bound to the Galactic plane. 
\end{abstract}

\section{Introduction}
\label{section1}

Galactic open clusters (OCs) have a wide age range, from 0 to almost 10 Gyr, and they are spread throughout the Galactic disc; therefore, they are widely used to characterize the properties of the Galactic disc, such as the morphology of the spiral arms of the Milky Way (MW) \citep{Bonatto_2006, VanDenBergh_2006,Vazquez_2008}, the stellar metallicity gradient (e.g., \citealt{Cunha_2016,Jacobson_2016,Frinchaboy_2013,Geisler_1997,Janes_1979}), the age-metallicity relation in the Galactic disc \citep{Magrini_2009,Salaris_2004,Carraro_1994,Carraro_1998}, and the Galactic disc star formation history \citep{DeLaFuente_2004}. OCs are thus crucial in developing a more comprehensive understanding of the Galactic disc.

OCs are generally considered to be archetypal examples of a simple stellar population \citep{Deng_2007}, because individual member stars of each OC are essentially homogeneous, both in age, dynamically (similar radial velocities and proper motions) and chemically (similar chemical patterns), greatly facilitating our ability to derive global cluster parameters from studying limited samples of stars. However, possible small inhomogeneous chemical patterns in OCs have been recently suggested, though only at the 0.02 dex level (e.g. Hyades; \citealt{Liu_2016}).

IC 166 ($l=130.071^{\circ}$, $b=-0.189^{\circ}$) is an intermediate-age OC ($\sim$1.0 Gyr, \citealt{Vallenari_2000,Subramaniam_2007}) located in the outer part of the Galactic disc (R$_{GC}\approx$13 kpc). Previous literature studies of this cluster used mainly photometric and low-resolution spectroscopic data. Detailed photometric studies were carried out by \citet{Subramaniam_2007,Vallenari_2000} and \citet{Burkhead_1969} in order to estimate its age, extinction, and distance. In addition, \citet{Dias_2014,Dias_2002,Loktin_2003} and \citet{Twarog_1997} have derived proper motions in the IC 166 field. \citet{Friel_1993} and \citet{Friel_1989} have estimated the radial velocity and metallicity of IC 166 from low-resolution spectroscopic data. In this work, we will for the first time provide an extensive, detailed investigation of its chemical abundances as well as its orbital parameters.

OCs are continuously influenced by destructive effects such as (1) evaporation \citep{Loyola_2013}, where some members reach the escape velocity after intracluster stellar encounters with other members, and/or via interaction with the Galactic tidal field, and (2) close encounters with giant interstellar clouds \citep{Gieles_2016}. Interactions with giant molecular clouds along their orbit in the Galactic disc have a high probability to eventually disrupt star clusters \citep{Lamers_2005,Gieles_2006,Lamers_2006}. These effects can lead to the dissolution of a typical open cluster in $\sim$10$^{8}$ years \citep{Friel_2013}. Thus, intermediate-age and old OCs ($\geq$1.0 Gyr) are rare by nature and are of great interest \citep{Friel_2014,Donati_2014,Magrini_2015,Tang_2017}. 
As these effects are generally less severe in the outer disc, OCs there have a higher chance of survival, providing a great opportunity to study this part of the Galaxy both chemically and dynamically. Moreover, IC 166 is located close to the region where a break in the metallicity gradient is suggested (between 10 kpc and 13 kpc from the Galactic center; \citealt{Frinchaboy_2013,Yong_2012,Reddy_2016}).  Accurate determination of the cluster's metallicity is helpful to constrain the nature of this possible break.

Large scale multi-object spectroscopic surveys, such as the Apache Point Observatory Galactic Evolution Experiment (APOGEE: \citealt{Majeswky_2017}) provide a unique opportunity to study a wide gamut of light-/heavy-elements in the H-band in hundreds of thousands of stars in a homogeneous way \citep{Gracia_2016,Hasselquist_2016,Cunha_2017}. In this work, we provide an independent abundance determination of several chemical species in the open cluster IC 166 using the Brussels Automatic Code for Characterizing High accUracy Spectra (BACCHUS: \citealt{Masseron_2016}), and compare them with the Apogee Stellar Parameter and Chemical Abundances Pipeline (ASPCAP: \citealt{Gracia_2016}). 

This paper is organized as follows.  Cluster membership selection is described in Section \ref{sec:TargetSelection}. In Section \ref{sec:Stellar_Parameters} we determine the atmospheric parameters for our selected members. In Section \ref{sec:sec4} we present our derived chemical abundances. A detailed description of the orbital elements is given in Section \ref{sec:dynamic}. We present our conclusions in Section \ref{sec:sec7}.

\section{Target Selection}
\label{sec:TargetSelection}

The Apache Point Observatory Galactic Evolution Experiment \citep[APOGEE:][]{Majeswky_2017} is one of the projects operating as part of the Sloan Digital Sky Survey IV \citep{Abolfathi_2017, Blanton2017}, aiming to characterize the Milky Way Galaxy's formation and evolution through a precise, systematic and large scale kinematic and chemical study. The APOGEE instrument is a near-infrared ($\lambda=1.51-1.70$ $\mu$m) high resolution ($R \approx 22,500$) multi-object spectrograph \citep{Wilson_2012} mounted at the SDSS 2.5 m telescope \citep{Gunn_2006}, with a copy now operating in the South at Las Campanas Observatory --the  2.5-meter Ir\'en\'ee du Pont telescope. The APOGEE survey has observed more than 270,000 stars across all of the main components of the Milky Way \citep{Zasowski_2013, Zasowski_2017}, achieving a typical spectral signal to noise ratio (S/N) $>$100 per pixel. The latest data release \citep[DR14:][]{Abolfathi_2017} includes all of the APOGEE-1 data and APOGEE data taken between July 2014 and July 2016. A number of candidate member stars of the open cluster IC 166 were observed by the APOGEE survey, and their spectra were released for the first time as part of the DR14 \citep{Abolfathi_2017}.

\begin{figure}
\hspace{-1.5cm}
\includegraphics[scale=0.55]{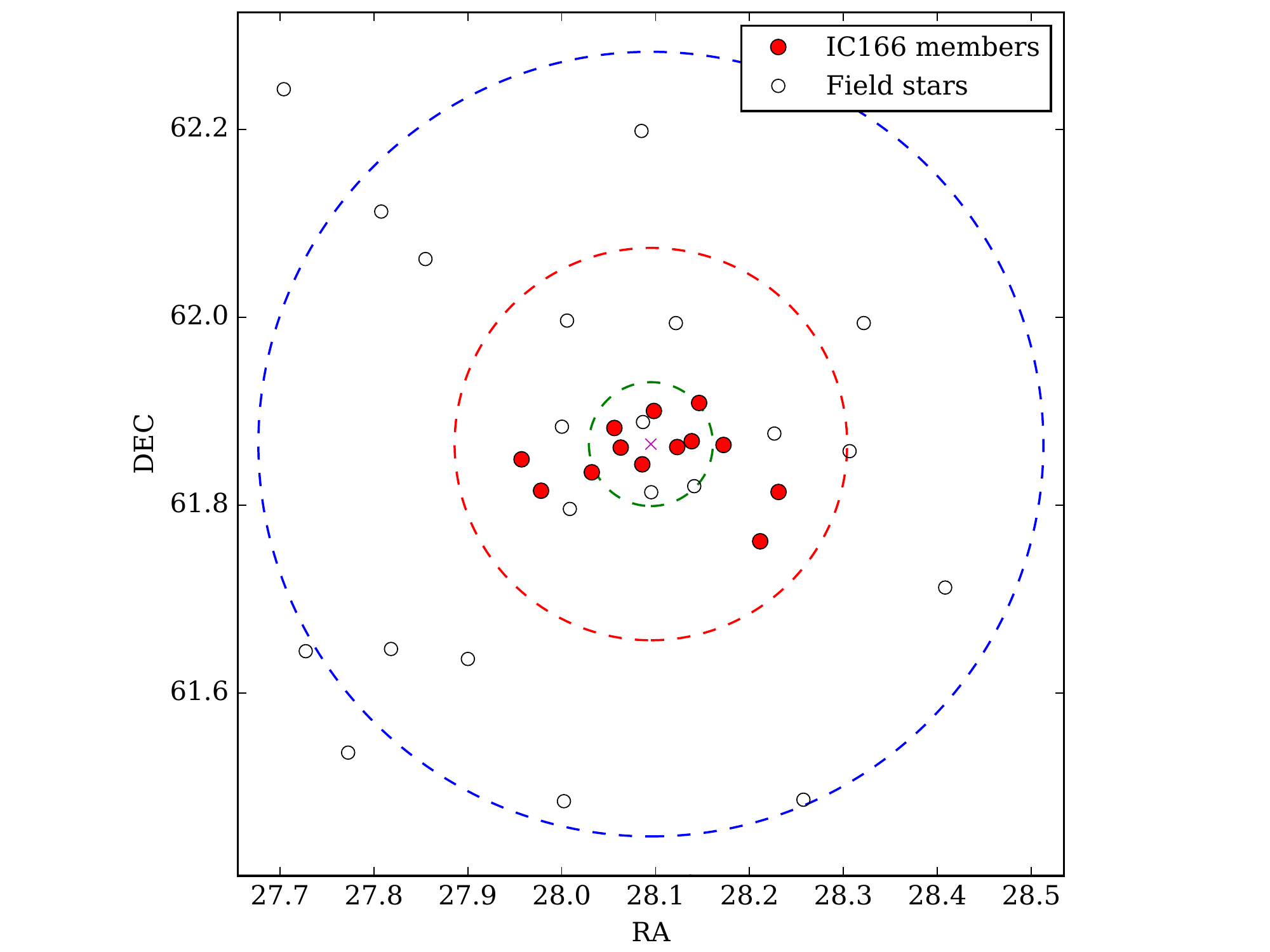}
    \caption{On-sky distribution of the 13 highest likelihood cluster members analyzed in this work (red symbols) and within 17.6 arcmin (half of the tidal radius) of the center (red dashed line). The inner "x" symbol is the center of the cluster. Indicated with black open circles are field stars that were also observed by APOGEE. The large blue dashed circle shows the tidal radius of the cluster (35.19 arcmin), while the inner green dashed circle shows the core radius of the cluster.}
    \label{fig:spatial_distribution}
\end{figure}

\begin{figure}
\centering
\includegraphics[width=2\linewidth]{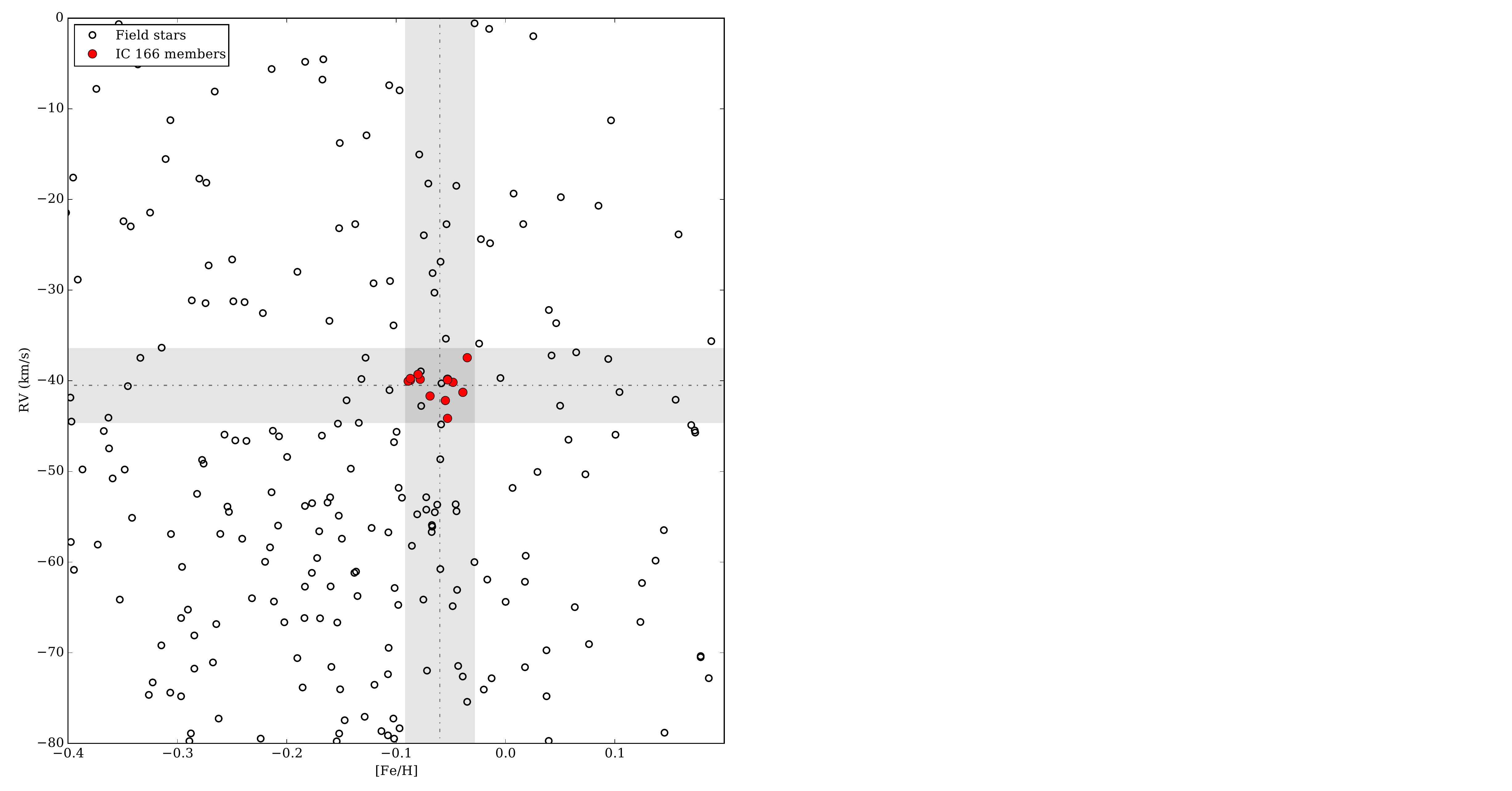}
    \caption{The APOGEE/DR14 RV vs. metallicity of stars in the field of the cluster (gray open circles) and our final sample (red dots). The gray regions show the upper and lower limits for the membership selection described in the text. The dotted lines show the mean RV and [Fe/H] of our final sample.}
    \label{fig:radial_velocity}
\end{figure}

\begin{figure}
\centering
\includegraphics[scale=0.5]{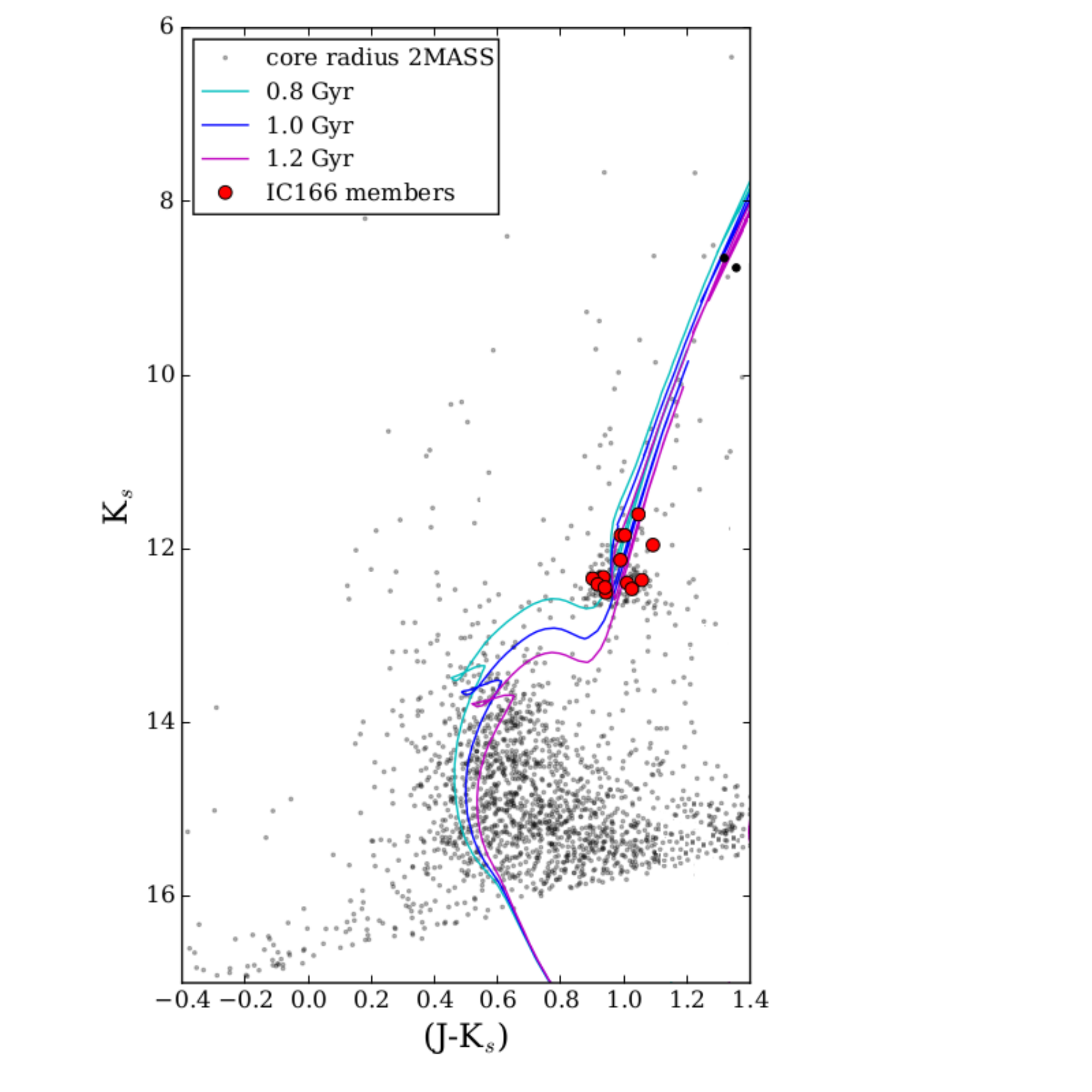}
    \caption{CMD of IC 166 using J and K$_{s}$ magnitudes. Small gray points represent stars observed by 2MASS inside of the $r_{core}$. Red dots represent our potential members observed by APOGEE and black dots the two stars not passing our high quality cuts (see text). Isochrones for 0.8 Gyr (sky-blue line), 1.0 Gyr (blue line) and 1.2 Gyr (magenta line) from PARSEC are also plotted.}
    \label{fig:cmd}
\end{figure}

We selected a sample of potential stellar members for IC 166 using the following high quality control cuts:

\begin{itemize}
\item[1.]{Spatial Location:} We focus on stars that are located inside half of the tidal radius ($r_{t}/2$), where $r_{t} = $ 35.19 $\pm$ 6.10 pc \citep{Kharchenko_2012}. This can minimize Galactic foreground stars. Figure \ref{fig:spatial_distribution} shows the spatial distribution of 21 highest likelihood cluster members inside half of the tidal radius, highlighted with red dots, for our final sample of likely cluster members. Stars with projected distances from the center larger than half of the tidal radius were removed, in order to obtain a cleaner sample, relatively uncontaminated by disc stars.

\item[2.]{Radial Velocity and Metallicity:} We further selected member stars using their radial velocities (RVs). Figure \ref{fig:radial_velocity} shows the RV versus [Fe/H] distribution of the stars in the APOGEE observation field of IC 166. Clearly, twenty out of twenty-one likely cluster members that we selected using only spatial information show a RV peak around $-40$ km s$^{-1}$, except one with much lower RV ($\approx -96$  km s$^{-1}$). The other 20 cluster members show a mean RV of $-40.50 \pm 1.66$ km s$^{-1}$. Applying a 3$\sigma$ limit, we excluded stars outside of $-40.50 \pm 3\times 1.66$ km s$^{-1}$ (gray region in Figure \ref{fig:radial_velocity}). Twenty stars were selected as likely members. After the spatial location and RV selection, their membership status is further scrutinized by filtering out all stars failing to meet the metallicity criteria. We adopt the calibrated metallicity from DR14 APOGEE/ASPCAP as a first guess in order to derive a cleaner sample of cluster stars. We identified a metallicity peak at $-$0.06 dex; thus stars with metallicities differing by more than 0.03 dex from this mean were removed. Fifteen stars were left as likely members.

\item[3.]{CMD Location:} The left panel of Figure \ref{fig:cmd} shows the 2MASS (K$_{\rm s}$, J-K$_{\rm s}$) Color-Magnitude diagram, for all stars lying inside one half of the tidal radius. 
Our selected APOGEE sample clearly lies near the red clump, consistent with the red clump observed in the $T_{\rm eff}$ vs. log(g) plane (right panel of Figure \ref{fig:cmd}). Interestingly, \citet{Vallenari_2000} also reported a clear red clump in IC 166, but did not find evidence of RGB stars. Two out of the fifteen stars selected previously were located away from the red clump of IC 166. 
These stars were also removed from further consideration, although isochrones indicate they could well be upper RGB members. The isochrones shown in Figure \ref{fig:cmd} were selected from PARSEC \citep{Bressan_2012} for [Fe/H] =-0.06 dex  and ages (0.8, 1.0 and 1.2 Gyr; \citealt{Vallenari_2000,Subramaniam_2007}) to match the metallicity and age reported for this cluster. 
The candidates are in good agreement with the selected isochrones. The PARSEC isochrones used have been fitted by eye to the luminosity and colour of the red clump stars. 
There is a small discrepancy in the location of de-reddened red clump stars found using the optical photometry and the T$_{eff}$ vs. log(g) diagram.

\end{itemize}

Lastly, we examine the newly-measured proper motions from \textit{Gaia} DR2 \citep{Gaia2018, Lindegren2018} of the APOGEE/IC 166 candidates. Figure \ref{fig:proper_motion} shows the proper motion diagram for IC 166. The dashed lines show the estimated mean proper motion value for IC 166. \textit{Gaia} DR2 reveals that the selected stars in this study exhibit similar proper motions to each other, with a relatively small spread ($<$ 0.2 mas yr$^{-1}$; see Figure \ref{fig:proper_motion}), which are good enough for a precise orbit predictions of IC 166.

\begin{figure}
\hspace{-1.cm}
\includegraphics[width=1.2\linewidth]{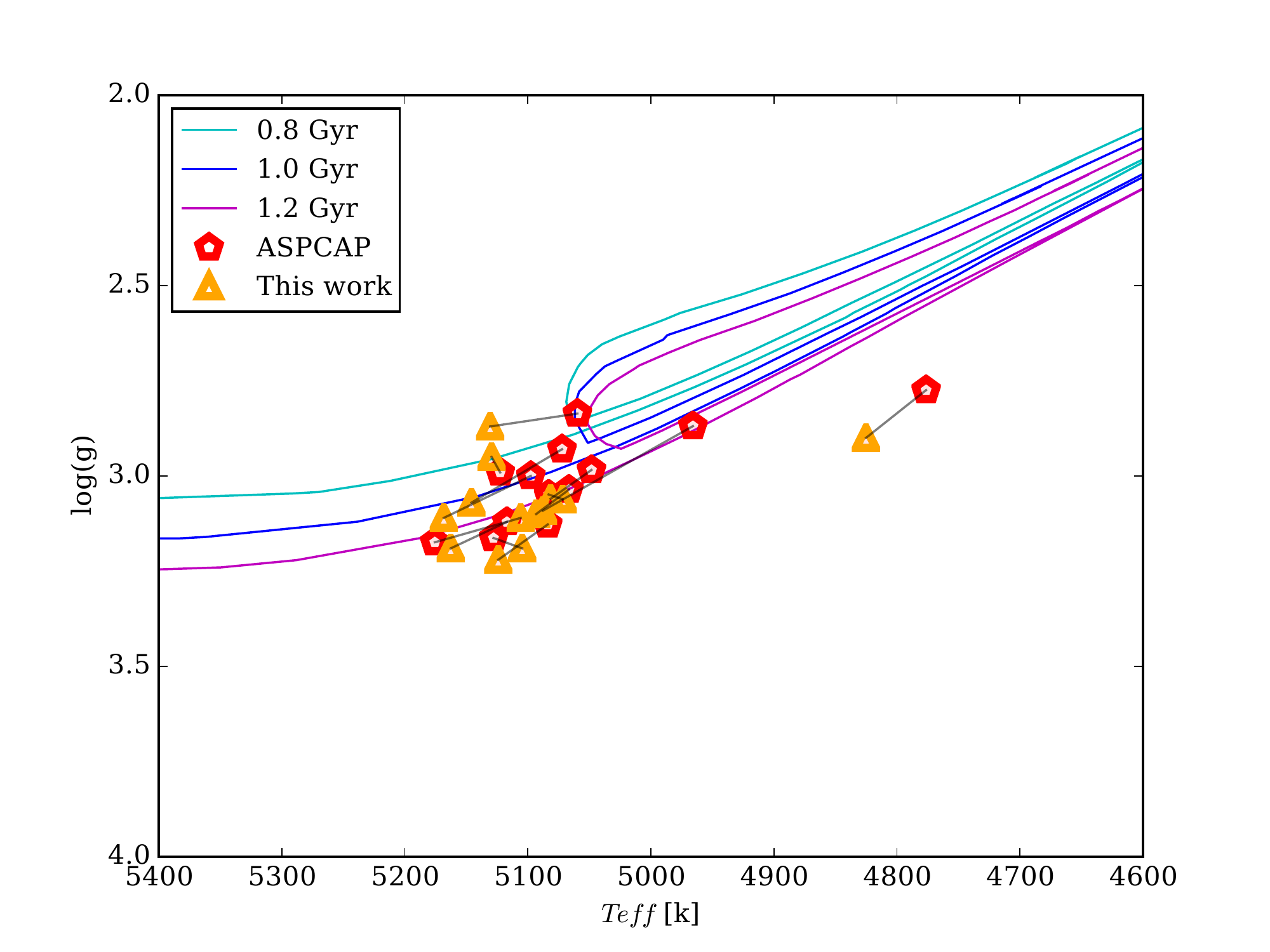}
    \caption{Log(g)-$T_{\rm eff}$ plane: Stellar parameters from ASPCAP and this work are represented with red pentagons and orange triangles, respective. Isochrones follow the same description as the Figure \ref{fig:cmd}. Black lines show which points refer to the same stars.}
    \label{fig:logg_teff}
\end{figure}

\begin{figure}
\hspace{-0.5cm}
\includegraphics[scale=0.50]{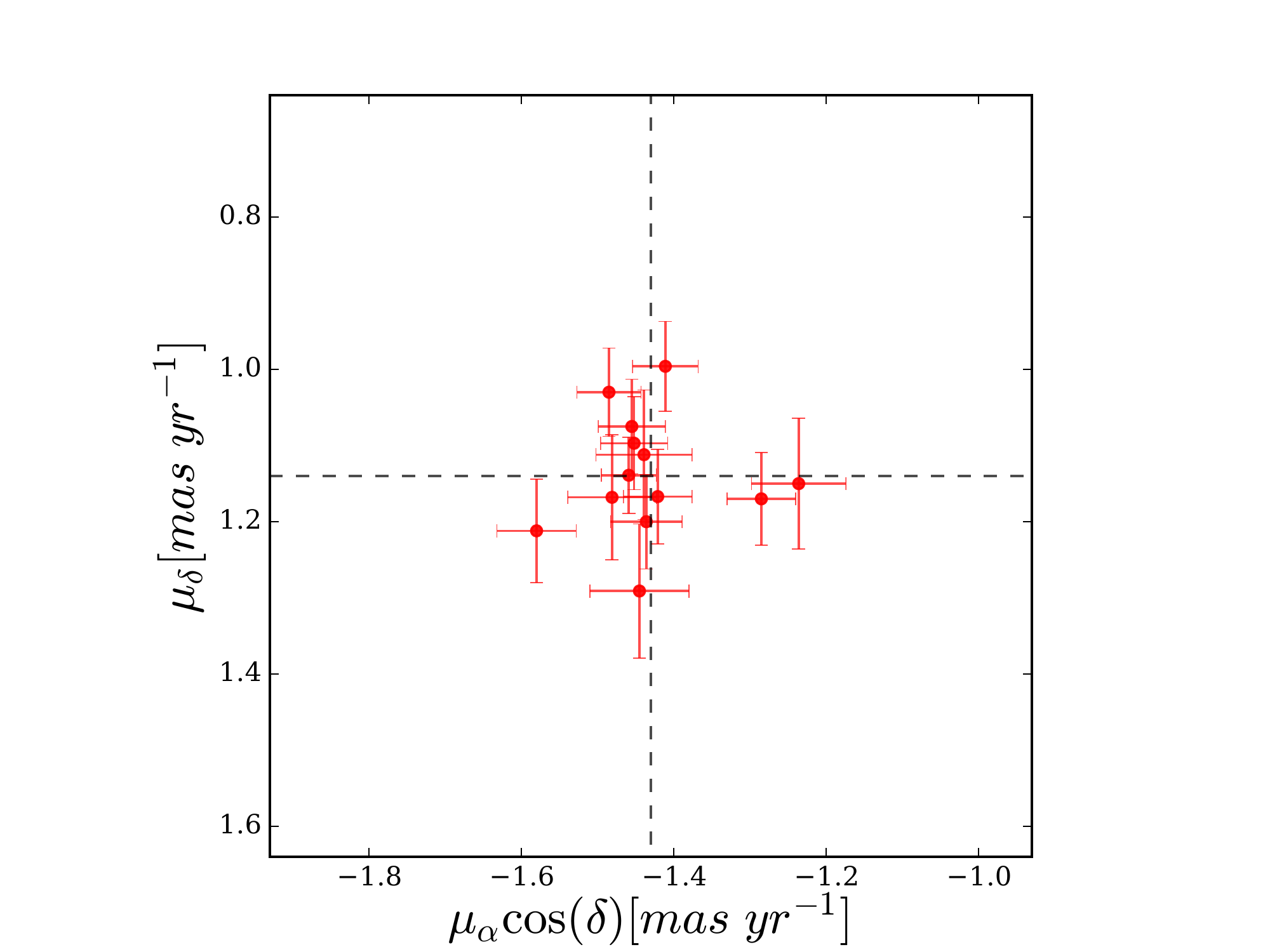}
    \caption{Proper motion diagram for the stars selected as members of IC 166 from \textit{Gaia} DR2. The dashed lines are the mean proper motion estimated for IC 166 (see text).}
    \label{fig:proper_motion}
\end{figure}

Table \ref{tab:datos_targets} shows the basic parameters of the stars which satisfy all the criteria previously mentioned, where raw $T_{\rm eff}$ and log(g) have been considered. These 13 stars will be considered as likely members of IC 166 and constitute our final cluster sample. 

\begin{table*}
\centering 
\caption{Summary table of likely members of IC 166.} 
\label{tab:datos_targets}
\begin{tabular}{ccccccc}
\hline
\hline
Apogee ID          & Tag       &  RA       & DEC       &   J      & K      &  \% $^{a}$ \\ \hline
2M01514975+6150556 & Star \#1  & 27.957296 & 61.848778 &   13.417 & 12.360  &  ...  \\ 
2M01515473+6148552 & Star \#2  & 27.978044 & 61.815334 &   13.403 & 12.393 &   95   \\ 
2M01520770+6150058 & Star \#3  & 28.032106 & 61.834946 &   13.446 & 12.503 &   96   \\ 
2M01521347+6152558 & Star \#4  & 28.056156 & 61.882183 &   13.487 & 12.462 &   ...  \\ 
2M01521509+6151407 & Star \#5  & 28.062883 & 61.861309 &   13.118 & 12.129 &   84   \\ 
2M01522060+6150364 & Star \#6  & 28.085842 & 61.843445 &   12.835 & 11.845 &   95   \\ 
2M01522357+6154011 & Star \#7  & 28.098241 & 61.900307 &   13.262 & 12.327 &   96   \\ 
2M01522953+6151427 & Star \#8  & 28.123055 & 61.861885 &   12.649 & 11.603 &   ...  \\ 
2M01523324+6152050 & Star \#9  & 28.138523 & 61.868073 &   13.244 & 12.343 &   63   \\ 
2M01523513+6154318 & Star \#10 & 28.146393 & 61.908844 &   13.326 & 12.409 &   ...  \\ 
2M01524136+6151507 & Star \#11 & 28.172348 & 61.864094 &   13.385 & 12.445 & 93   \\ 
2M01525074+6145411 & Star \#12 & 28.211422 & 61.76144  &   13.048 & 11.956 &   ...  \\
2M01525543+6148504 & Star \#13 & 28.230962 & 61.814007 &   12.847 & 11.844 &   ...  \\ 
\hline
\hline 
\end{tabular}
\\
\hspace{4cm}
\raggedright{$^a$ Membership probability from \citet{Dias_2014}}
\end{table*}

\section{Atmospheric parameters and abundance determinations}
\label{sec:Stellar_Parameters}

For the stars observed with APOGEE and identified as members in \S\ref{sec:TargetSelection}, atmospheric parameters (T$_{\rm eff}$, log g, [M/H] and $\xi$) were determined using the code FERRE \citep{AllendePrieto_2006} that compares theoretical spectra computed from MARCS atmosphere models \citep{Gustafsson_2008,Zamora_2015} using the entire wavelength range, and minimizes the difference with the observed spectrum via a $\chi^2$ minimization. Our synthetic spectra were based on 1D Local Thermodynamic Equilibrium (LTE) model atmospheres calculated with MARCS \citep{Gustafsson_2008}. The derived atmospheric parameters are listed in Table \ref{tab:stellar_parameters}.

It is important to note that we chose not to estimate the T${\rm eff}$ values from any empirical color-temperature relation; this would be highly uncertain due to relatively large and likely differential reddening along the line-of-sight to IC 166, $E(B-V)\approx$ 0.80 \citep{Subramaniam_2007}.

Figure \ref{fig:logg_teff}  displays the main stellar parameters determined from FERRE/MARCS against those computed from ASPCAP/KURUCZ (raw values), overplotted on the PARSEC isochrones \citep{Bressan_2012} with ages of 0.8, 1.0 and 1.2 Gyr. We notice that the raw (not post-calibrated) stellar parameters obtained via ASPCAP/KURUCZ are in fairly good agreement with the stellar parameters derived in this study using FERRE/MARCS. After deriving the stellar parameters we used the code BACCHUS \citep[see][]{Masseron_2016, Hawkins_2016} to fit the spectral features of the atomic lines for up to eight chemical elements (Fe, Mg, Al, Si, Ca, Ti, K, and Mn). We did not analyze OH, CN, and CO, because these molecular lines are weak in the typical range of T$_{\rm eff}$ and metallicity for the stars studied in this work, and such an analysis would lead to unreliable abundance results for carbon, nitrogen and oxygen. The line list used in this work is the latest internal DR14 atomic/molecular line list (linelist.20150714: Holtzmman et al. in preparation). For each atomic line, the abundance determination proceeded in the same fashion as described in \citet{Hawkins_2016}, i.e., we computed spectrum synthesis, using the full set of atomic lines to find the local continuum level via a linear fit; the local S/N was estimated and the abundances were then determined by comparing the observed spectrum with the set of convolved synthetic spectra for different abundances. The BACCHUS code determines line-by-line abundances via four different approaches: (i) line-profile fitting; (ii) core line intensity comparison; (iii) global goodness-of-fit estimate ($\chi^2$); and (iv) equivalent width comparison, with each diagnostic yielding validation flags used to reject or accept a line, keeping the best fit abundance \cite[see, e.g.,][]{Hawkins_2016}. Following the suggestion by \cite{Hawkins_2016,FT_2018b} we adopted the $\chi^{2}$ diagnostic as the most robust abundance determination. The selected atomic lines were then visually inspected to ensure that the spectral fits were adequate. The spectral regions used in our analysis are listed in Table \ref{tab:result_ferre}.

\begin{table*}
\centering
\caption{Stellar parameters obtained from FERRE/MARCS.}
\label{tab:stellar_parameters}
\begin{tabular}{ccccccccc}
\hline 
\hline
         \multicolumn{5}{c}{This work} & \multicolumn{4}{c}{ASPCAP}\ \\ 
\hline
ID          &  T${\rm eff}$        & log(g) &   [Fe/H] &  $\xi$ &   T${\rm eff}$      & log(g) & [Fe/H]   &  $\xi$ \\ \hline
 star \#1   &  5070        & 3.06   &   -0.01  & 1.32         &   5085      &  3.05  & -0.05    &  1.70        \\
 star \#2   &  5080        & 3.06   &   -0.08  & 1.25         &   5050      &  3.00  & -0.08    &  1.50        \\ 
 star \#3   &  5130        & 2.95   &   -0.05  & 0.93         &   5120      &  3.00  & -0.09    &  1.10        \\ 
 star \#4   &  5095        & 3.10   &   -0.05  & 1.16         &   5065      &  3.05  & -0.05    &  1.60        \\ 
 star \#5   &  5145        & 3.07   &   -0.04  & 1.55         &   5070      &  2.95  & -0.05    &  1.70        \\ 
 star \#6   &  5105        & 3.19   &   -0.05  & 1.06         &   5130      &  3.15  & -0.03    &  1.50        \\
 star \#7   &  5170        & 3.11   &   -0.10  & 1.42         &   5100      &  3.00  & -0.09    &  1.60        \\ 
 star \#8   &  4825        & 2.90   &   -0.07  & 1.19         &   4775      &  2.75  & -0.04    &  1.50        \\ 
 star \#9   &  5105        & 3.11   &   -0.11  & 1.24         &   5175      &  3.15  & -0.05    &  1.60        \\ 
 star \#10  &  5165        & 3.19   &   -0.10  & 1.28         &   5115      &  3.10  & -0.08    &  1.70        \\ 
 star \#11  &  5125        & 3.22   &   -0.06  & 1.08         &   5085      &  3.15  & -0.05    &  1.45        \\ 
 star \#12  &  5130        & 2.87   &   -0.06  & 1.69         &   5060      &  2.85  & -0.09    &  1.70        \\ 
 star \#13  &  5090        & 3.09   &   -0.05  & 1.42         &   4965      &  2.85  & -0.07    &  1.70        \\
 \hline
 \hline
\end{tabular}
\end{table*}

\begin{table*}
	
	\caption{IC 166 Data Sample from \textit{Gaia} DR2 and APOGEE.}
	\label{tab:orbits}
	\centering
		\begin{tabular}{ccccccc}
			\hline 
			\hline
		  APOGEEID           &     $\alpha$  &   $\delta$    &         Parallax         &       Radial velocity  &    $\mu_{\alpha} $  &   $\mu_{\delta} $  \\
		  &      [J2000]    &      [J2000]    &          [mas]           &         [km s$^{-1}$]  &        mas yr$^{-1}$      &    mas yr$^{-1}$ \\
		  \hline
		  \hline
		  2M01514975+6150556 &  01:51:49.75  &  +61:50:55.6  &   0.188 $\pm$  0.055     &    -39.817 $\pm$ 0.367 &  -1.439 $\pm$ 0.063 &  1.112 $\pm$ 0.085 \\
		  2M01515473+6148552 &  01:51:54.73  &  +61:48:55.2  &   0.177 $\pm$  0.054     &    -39.834 $\pm$ 0.326 &  -1.481 $\pm$ 0.058 &  1.168 $\pm$ 0.082 \\
		  2M01520770+6150058 &  01:52:07.71  &  +61:50:05.8  &   0.305 $\pm$  0.056     &    -39.951 $\pm$ 0.355 &  -1.236 $\pm$ 0.062 &  1.150 $\pm$ 0.086 \\
		  2M01521347+6152558 &  01:52:13.48  &  +61:52:55.9  &   0.227 $\pm$  0.059     &    -44.142 $\pm$ 0.688 &  -1.445 $\pm$ 0.065 &  1.291 $\pm$ 0.088 \\
		  2M01521509+6151407 &  01:52:15.09  &  +61:51:40.7  &   0.146 $\pm$  0.040     &    -40.167 $\pm$ 0.211 &  -1.455 $\pm$ 0.044 &  1.075 $\pm$ 0.062 \\
		  2M01522060+6150364 &  01:52:20.60  &  +61:50:36.4  &   0.125 $\pm$  0.044     &    -37.449 $\pm$ 0.092 &  -1.436 $\pm$ 0.047 &  1.200 $\pm$ 0.062 \\
		  2M01522357+6154011 &  01:52:23.58  &  +61:54:01.1  &   0.226 $\pm$  0.040     &    -40.035 $\pm$ 0.192 &  -1.452 $\pm$ 0.044 &  1.097 $\pm$ 0.061 \\
		  2M01522953+6151427 &  01:52:29.53  &  +61:51:42.8  &   0.181 $\pm$  0.034     &    -41.271 $\pm$ 0.133 &  -1.459 $\pm$ 0.036 &  1.139 $\pm$ 0.050 \\
		  2M01523324+6152050 &  01:52:33.25  &  +61:52:05.1  &   0.175 $\pm$  0.040     &    -42.174 $\pm$ 0.110 &  -1.285 $\pm$ 0.045 &  1.170 $\pm$ 0.061 \\
		  2M01523513+6154318 &  01:52:35.13  &  +61:54:31.8  &   0.092 $\pm$  0.041     &    -39.288 $\pm$ 0.093 &  -1.421 $\pm$ 0.045 &  1.167 $\pm$ 0.062 \\
		  2M01524136+6151507 &  01:52:41.36  &  +61:51:50.7  &   0.161 $\pm$  0.048     &    -41.968 $\pm$ 0.033 &  -1.580 $\pm$ 0.052 &  1.212 $\pm$ 0.068 \\
		  2M01525074+6145411 &  01:52:50.74  &  +61:45:41.2  &   0.175 $\pm$  0.038     &    -39.743 $\pm$ 0.147 &  -1.485 $\pm$ 0.042 &  1.030 $\pm$ 0.058 \\
		  2M01525543+6148504 &  01:52:55.43  &  +61:48:50.4  &   0.224 $\pm$  0.039     &    -41.678 $\pm$ 0.176 &  -1.411 $\pm$ 0.043 &  0.996 $\pm$ 0.059 \\
			\hline
			\hline
		\end{tabular}
\end{table*}

In this study, we derived the abundances of the elements Mg, Si, Ca and Ti ($\alpha$-elements);  Al and K (light odd-Z elements); and Mn and Fe (iron-peak elements). Our abundances were scaled relative to solar abundances \citep{Asplund_2005}, in order to provide a direct comparison with the ASPCAP determinations.\\

\section{Results and discussion}
\label{sec:sec4}
\subsection{Chemical abundances from BACCHUS vs. ASPCAP}

As mentioned above, we derived chemical abundances manually for our sample stars using the BACCHUS code and using the stellar parameters obtained with FERRE/MARCS.
Line-by-line abundance determinations were done for each element for each studied star  (Appendix \ref{sec:apendice}).  
The wavelengths of the selected transitions (in air wavelength) for each element are listed in the second column of Table \ref{tab:result_ferre}. Both abundances A(X) and the solar scaled abundances are given. The `...' symbol is used to indicate that it was not possible to measure a line due to effects such as saturation, weak line, noise, or blending.

Fe, Mg, and Si are the elements having both stronger and more numerous lines in the APOGEE spectra, with 9, 3 and 14 measured lines, respectively. For potassium, we could only identify one K I line in a few of the stars, which is also the case for Ti. We decided to eliminate from further study the Ti abundances due to large uncertainties. In addition, the derived K abundances should be used with caution.

Table \ref{tab:results} shows the average abundances of Mg, Si, Ca, Al, K, Mn and Fe, from our manual analysis against the ASPCAP determinations. These results will be compared below:

\begin{itemize}
\item Mg: The mean [Mg/Fe]$_{\rm our}$ abundance ratio is systematically lower (by $\sim$ -0.19 dex) when compared to [Mg/Fe]$_{ASPCAP}$, but shows a dispersion that is comparable to [Mg/Fe]$_{ASPCAP}$. Magnesium is, by far, the element most affected by the change of stellar parameters.

\item Si: The mean [Si/Fe]$_{\rm our}$ abundance ratio is very similar to that of ASPCAP, ours being just slightly lower (by 0.01 dex) than ASPCAP. However,

[Si/Fe]$_{ASPCAP}$ shows larger scatter when compared to our results. 

\item Ca: [Ca/Fe]$_{\rm our}$ has a small offset of 0.05 dex in the mean abundance when compared with ASPCAP, with both sets of resuts finding the same scatter of 0.04 dex. 

\item Al: The mean [Al/Fe]$_{ASPCAP}$ abundance is offset by 0.07 dex when compared to our mean [Al/Fe]$_{\rm our}$ abundance ratio. Most importantly, [Al/Fe]$_{ASPCAP}$ show a large dispersion (0.37 dex), which is not consistent with the homogeneity expected in OCs. This is likely related to night-sky OH contamination of some of the 3 stronger red lines of Al I that are not properly accounted for in the automatic pipeline analysis. The result of this improper treatment in ASPCAP would be increased weight in the final Al abundances given to the very weak Al I blue lines at $\lambda$ 15956.675 \AA, and $\lambda$ 15968.287 \AA. Our abundance results have a very small scatter of 0.05 dex, which is similar to what is found for the other studied elements.

\item K: The mean [K/Fe]$_{\rm our}$ abundance ratio is slightly higher than the mean [K/Fe]$_{ASPCAP}$, but the values are in agreement within the uncertainties. Because we could only measure one line for K I in most of the stars, the K abundance results should be taken with caution.

\item Mn: [Mn/Fe]$_{\rm our}$ are in agreement with [Mn/Fe]$_{ASPCAP}$. All of them have abundances close to solar.

\item Fe: The mean [Fe/H]$_{\rm our}$ abundance ratio is slightly lower (by 0.02 dex; [Fe/H]$_{\rm our}=-0.08\pm0.05$) than [Fe/H]$_{ASPCAP}$. ASPCAP finds a very small scatter in the iron abundances in this cluster, while our sigma is 0.05 dex, compatible with what is found for the other studied  elements.  

\end{itemize}

In general, there is good agreement between the mean abundances obtained manually in this work and the ASPCAP values with comparable dispersion, except for Mg. For Al it is clear that there is a problem with the ASPCAP abundances in this cluster; these issues will be corrected in DR15.

\begin{table}
\centering
\caption{Mean chemical abundances and dispersions
for thirteen likely members of IC 166.}
\label{tab:results}
\begin{tabular}{ccc}
\hline 
\hline
Element &  This work      &  ASPCAP        \\ \hline
Mg      &  -0.18$\pm$0.04 &  0.01$\pm$0.04 \\ 
Si      &   0.06$\pm$0.02 &  0.07$\pm$0.06 \\ 
Ca      &  -0.05$\pm$0.04 &  0.00$\pm$0.04 \\ 
Al      &   0.11$\pm$0.05 &  0.05$\pm$0.37 \\ 
K       &   0.00$\pm$0.08 & -0.04$\pm$0.08 \\ 
Mn      &  -0.02$\pm$0.03 &  0.00$\pm$0.03 \\ 
Fe      &  -0.08$\pm$0.05 & -0.06$\pm$0.02 \\ 
\hline
\hline
\end{tabular}
\end{table}

\subsection{Uncertainties}

The uncertainties of chemical abundances are estimated by perturbing the input stellar parameters. We chose star 2 as a representative of our sample stars. We vary each stellar parameter individually according to it own uncertainty ($\Delta$T$_{\rm eff}$=+50 K, $\Delta$log(g)=+0.20 dex, $\Delta$[Fe/H]=+0.20 dex, $\Delta \xi$=+0.20 km s$^{-1}$) in a similar way as described by \citet[][]{Souto_2016}, and measure the chemical abundances again. 

The differences in chemical abundances measured assuming perturbed stellar parameters and unperturbed ones are listed in Table \ref{tab:uncertainties}. Overall, the chemical abundance uncertainties caused by stellar parameter uncertainties are around 0.1 dex, with slightly larger uncertainties for Mg and K. Mg is mostly affected by variation of T$_{\rm eff}$ and log g, while K is mostly affected by variation of T$_{\rm eff}$ and [Fe/H].

\begin{table*}
\centering
\caption{Estimated uncertainties on abundances due to stellar parameter uncertainties.}
\label{tab:uncertainties}
\begin{tabular}{ccccccccc}
\hline \hline
Element	&	$\Delta$T${\rm eff}$	&	$\Delta$log(g)	&	$\Delta$[Fe/H]	&	$\Delta \xi$	&	$\sigma$	\\
	&	(+50 K)	&	(+0.20 dex)	&	(+0.20 dex)	&	(+0.20 km s$^{-1}$)	&	\\ \hline 
Mg	&	+0.06	&	-0.09	&	+0.01	&	-0.02	&	0.11	\\
Si	&	+0.03	&	-0.05	&	+0.02	&	-0.03	&	0.07	\\
Ca	&	+0.05	&	-0.02	&	-0.04	&	-0.02	&	0.07	\\
Al	&	+0.05	&	-0.07	&	+0.00	&	-0.02	&	0.09	\\
K	&	+0.06	&	-0.01	&	-0.15	&	-0.01	&	0.16	\\
Mn	&	+0.03	&	-0.01	&	-0.04	&	-0.02	&	0.05	\\
Fe	&	+0.05	&	-0.02	&	+0.02	&	-0.03	&	0.06	\\
\hline \hline
\end{tabular}
\end{table*}

 \subsection{Comparison with the literature}
Many studies have attempted to trace and understand the formation history and chemical evolution of the Galactic thin and thick discs, bulge and halo (e.g, \citealt{Bensby_2014,Battistini_2015,Nissen_2000}), aided by homogeneous and large data sets such as APOGEE \citep{Majeswky_2017}, Gaia-ESO \citep{Randich_2013,Gilmore_2012} and GALAH \citep{DeSilva_2015}.

\begin{figure*}
\centering
\includegraphics[width=\textwidth]{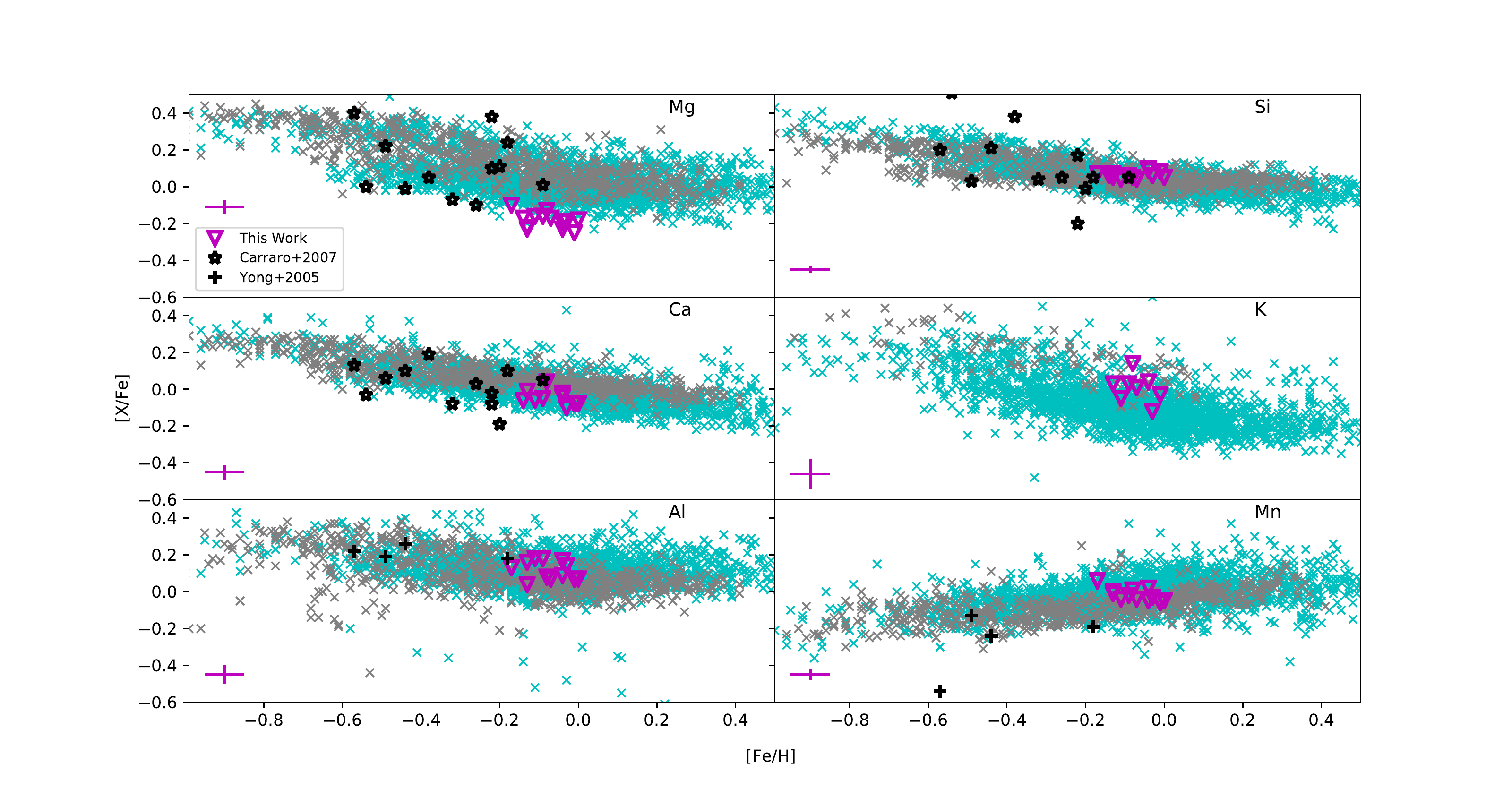}
    \caption{IC 166 results are compared with the  literature, cyan crosses for \citet{Hawkins_2016} results, while gray crosses for Ca, Mg, Si, and Al abundances from \citet{Bensby_2014}. The sources of the gray crosses for K and Mn are \citet{Nissen_2000} and \citet{Battistini_2015}, respectively. The Galactic anti-center OCs from \citet{Carraro_2007} and \citet{Yong_2005} are shown as black stars and black pluses, respectively.}
    \label{fig:trend1}
    \end{figure*}

To compare with our results (see Figure \ref{fig:trend1}), we have assembled (1) a sample of dwarf stars in the solar neighborhood from \citet{Bensby_2014} (gray crosses in Mg, Ca, Si and Al panels); (2) a sample of dwarf stars in the solar neighborhood from \citet{Battistini_2015} (gray crosses in the Mn panel); (3) a sample of F and G main-sequence stars of the disc taken from \citet{Nissen_2000} (gray crosses in the K panel); (4) a sample of cluster and field red giants and a small fraction of dwarf stars from APOGEE-Kepler Asteroseismology Collaboration (henceforth, APOKASC sample) which were re-analyzed using BACCHUS by \citet{Hawkins_2016} (cyan crosses in Mg, Si, Ca, K, Al and Mn panels); 
and (5) Galactic anti-center OCs from \citet{Carraro_2007} and \citet{Yong_2005} (black stars and black pluses, respectively). Finally, we show our results using FERRE/MARCS stellar parameters, which are represented with magenta triangles.

\begin{itemize}
\item $\alpha$-elements: Magnesium, silicon and calcium are generally considered as $\alpha$-elements, because they are formed by fusion involving $\alpha$-particles. These elements can be produced in large quantities by Type II supernovae \citep{Samland_1998}. $[\alpha$/Fe] decreases while metallicity increases after the onset of Type Ia SNe (e.g., \citealt{Bensby_2014} and \citealt{Hawkins_2016}). If we look closely, this trend is separated into two sequences, especially for Mg. Using the APOGEE data, \citet{Hayden_2015} found that the higher [$\alpha$/Fe], more metal-poor sequence is dominated by thick disc stars, while the lower [$\alpha$/Fe], more metal-rich sequence is dominated by thin disc stars. In general, our results follow the pattern formed by (thin disc) field stars at the metallicity of IC 166 members, except for Mg. However, similar to \citet{Hawkins_2016}, the magnesium abundances that we have found from our manual analysis are lower compared to \citet{Bensby_2014}, though they still agree within the uncertainties. \\

Galactic anti-center OCs from \citet{Carraro_2007} form trends of [$\alpha$/Fe$]-[$Fe/H] very similar to the field stars: [$\alpha$/Fe] decreases as [Fe/H] increases. It also appears that the scatter of [Mg/Fe] is larger than that of [Si/Fe] and [Ca/Fe], which also closely resembles field stars. IC 166 is one of the Galactic anti-center OCs with relatively high metallicity. Its $\alpha$-element abundances generally fit in the trends defined by other Galactic anti-center OCs, though with slightly lower Mg abundances.

\item Light odd-Z elements: Potassium is primarily the result of oxygen burning in massive stellar explosions \citep{Clayton_2007}, so it is related to  $\alpha-$element formation \citep{Zhang_2006}, expelled from Type II supernovae \citep{Samland_1998}. Although there is not much observational data for K, the available data indicate that its abundance increases as metallicity decreases. Figure \ref{fig:trend1} shows that the results from \citet{Nissen_2000} are shifted to higher abundances with respect to that of \citet{Hawkins_2016}, probably due to differences in the adopted solar abundances. 

Our results follow the expected trend for field stars at this metallicity.\\
Aluminum is formed during carbon burning in massive stars, mostly 
by the reactions between $^{26}$Mg and excess neutrons \citep{Clayton_2007}. The Al abundances may also be changed through the Mg-Al cycle at extremely high temperature, 
e.g., inside AGB stars \citep{Samland_1998,Arnould_1999}.  Literature values indicate that the Al abundance decreases as metallicity increases, and it stays relatively 
constant for metallicity greater than solar. The large dispersion found in the ASPCAP Al abundance results for IC 166 are not found in the literature for any open cluster, nor in our manual results. As discussed above this is due to problems in the ASPCAP analysis. Four Galactic anti-center OCs from \citet{Yong_2005}, together with IC 166 form a similar [Al/Fe$]-[$Fe/H] trend as field stars.

\item Iron-peak elements: Manganese is thought to form in explosive silicon burning \citep{Woosley_1995,Clayton_2007,Battistini_2015}. Significant amounts of Manganese are produced by both SN type II and SN type Ia \citep{Clayton_2007}. According to the observations, Mn closely follows Fe. Our results for Mn fall within the abundance distribution outlined by field stars at similar metallicity. Three Galactic anti-center OCs from \citet{Yong_2005}, together with IC 166 form a similar [Mn/Fe$]-[$Fe/H] trend as field stars. Exception is found for Be 31, where \citet{Yong_2005} suggested observations of additional members of Be 31 are required to confirm low [Mn/Fe] in all Be 31 cluster members.

\end{itemize}
To summarize, the results obtained in this study (using the BACCHUS code) are in good agreement with literature results about field giant/dwarf stars. The chemical abundances also verify that IC 166 is a typical anti-Galactic center OC, with relatively high metallicity among the others. 

\subsection{OC Metallicity trend around R$_{GC}$ of IC 166}
\label{sec:metalicity_gradient}

Studies of the Galactic radial metallicity gradient \citep{Friel_1995,Frinchaboy_2013,Jacobson_2016,Cunha_2016} are critical to understand the chemical evolution of the Galactic disc. Open clusters are one of the best tracers for this purpose, because they are located along the whole Galactic disc and they provide relatively easily measured chemical and kinematic properties. Most works agree that the metallicity decreases with increasing Galactic radius, at least for older OCs. However, the exact value of the metallicity gradient slope is still unclear \citep{Jacobson_2016,Cunha_2016}, nor the location of a possible break in the metallicity trend \citep{Yong_2012, Reddy_2016}. 

In this work, we analyze the high-resolution spectra of IC 166 stars, and derive a metallicity of [Fe/H$]=-0.08\pm0.05$ dex. Since IC 166 (R$_{GC}\approx $12.7 kpc) is located near the possible transition zone around R$_{GC}\approx 10-13$ kpc \citep{Frinchaboy_2013,Yong_2012, Reddy_2016}, it may be enlightening to compare our results to the other high-resolution chemical abundance analysis on OCs near this region. 
For example, at R$_{GC}\approx $10.5 kpc, \citet{Sales_2016} derived a metallicity of $-0.02\pm0.05$ dex for Tombaugh 1; \citet{Souto_2016} reported a metallicity of $-0.16\pm0.04 $ dex for NGC 2420 at R$_{GC}\approx $11 kpc. More strikingly, \citet{Reddy_2016} showed that the metallicities of OCs between $10-13$ kpc (including about 15 OCs) vary between 0 and $-0.4$ (their Figure 4). They suggested this region is the transition zone between thin disc OCs and thick disc OCs. Therefore, IC 166, Tombaugh 1, and NGC 2420 safely fit in the metallicity range defined by other OCs in this region. A discussion about the existence of this break requires a large number of OCs at different R$_{GC}$, which is certainly beyond the scope of this single OC concentrated work. Readers are referred to \citet{Yong_2012, Reddy_2016} for discussion about this topic.

\subsection{[\texorpdfstring{$\alpha$}//Fe] versus [Fe/H] }
As noted above, $\alpha$-elements are formed from reactions with $\alpha$-particles (He nuclei), which are active in Type II SNe. On the other hand, Fe is generated in SN Ia (although also, in smaller amounts, in SNe II); therefore [$\alpha$/Fe] is related to the ratio of Type II over Type Ia SNe that have enriched a particular star-forming environment. 

Because the main polluters of the ISM in the early stages of galaxy formation are Type II SNe, we see enhanced $\alpha$-element abundances at low metallicities. After $\sim 1$ Gyr, Type Ia SNe start to explode, generating a significant amount of iron-peak elements but insignificant amounts of $\alpha$-elements, and the iron-peak element fraction in the ISM increases quickly \citep{Bensby_2005}; [$\alpha$/Fe] decreases as the metallicity increases.

\begin{figure*}
\centering
\includegraphics[scale=0.75]{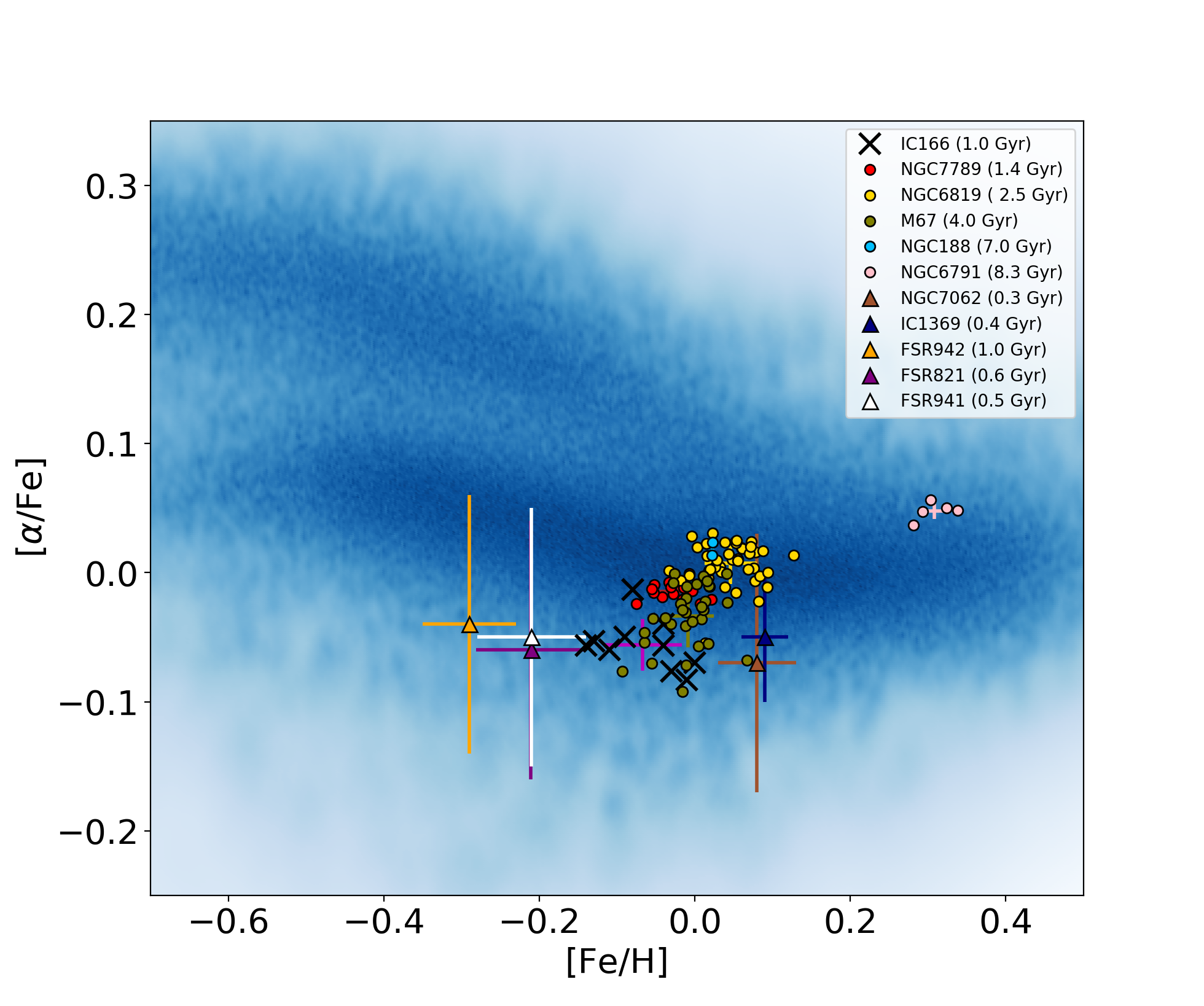}
    \caption{Density map for [$\alpha$/Fe] versus [Fe/H], illustrating the high and low $\alpha$-sequence formed by thick and thin disc stars observed by APOGEE DR14. Five open clusters studied by \citet{Linden_2017} (M67, NGC 7789, NGC 6819, NGC6791, NGC 188 in green, red, yellow, pink, and cyan dots, respectively) are also shown.  A sample of five new OCs (NGC 7062, IC1369, FSR 942, FSR 821, FSR 941 in brown, blue, orange, purlple and white triangles, respectively) studied by \citet{Frinchaboy_2013} poorly alpha-enriched. Stars of IC 166 are shown as black "x" symbols. The colored crosses show the mean abundances and the standard deviation for each cluster.
    	}
    \label{fig:alpha-trend}
\end{figure*}

Figure \ref{fig:alpha-trend} shows stars from APOGEE DR14 as gray dots. Results from ASPCAP DR13 for five OCs (M67, NGC 7789, NGC 6819, NGC 6791, NGC 188 in green, red, yellow, blue and pink dots, respectively) studied by \citet{Linden_2017} are added, and also the results from this study for IC 166 (purple dots). The $\alpha$-elements in this plot are an average of the elements Mg, Si and Ca. All of the IC 166 stars are close to the expected trend for thin disc stars (with a mean [$\alpha$/Fe]$\sim$-0.05), although our abundance results are slightly more scattered when compared to the results for the other clusters. 

IC 166 falls within the region of low $\alpha$-sequence. Thus the chemical signatures of IC 166, appear to follow the same abundance trends as thin disc field stars  (see Figure \ref{fig:alpha-trend}); very similar to other know disc OCs like NGC 7062, IC1369, FSR 942, FSR 821 and FSR 941 studied in \citet{Frinchaboy_2013}.

\section{The orbit of IC 166}
\label{sec:dynamic}

In order to estimate for the first time a probable Galactic orbit for IC 166, the positional information of IC 166, $(\alpha_{J2000}, \delta_{J2000}) = 01^{h} 46^{m},61^{\circ} 23'$, was combined with the newly-measured proper motions and parallaxes from Gaia DR2 \citep{Gaia2018, Lindegren2018} as well as with the existing line-of-sight velocities from the APOGEE survey. There were 13 stars in our sample, which were in the \textit{Gaia} DR2 catalogue and had a good parallax signal-to-noise ($\pi/\pi_{\omega} > 3$; see Table \ref{tab:orbits}). For the 13 members surveyed by APOGEE (for which the membership is most certain) we estimate the mean proper motion of IC 166 as ($\mu_{\alpha}, \mu_{\delta} $) $=$ (-1.429$\pm$0.083 , 1.139$\pm$0.075) mas yr$^{-1}$, a radial velocity of -40.58$\pm$1.59 km s$^{-1}$, and a median parallax, ($\langle \pi \rangle \pm \sigma_{\pi}$)$= (0.18466 \pm 0.05095)$  , distance of 5.415$\pm$1.494 kpc, our distance estimated from parallax tend to agree with the mean distance estimated from a Bayesian approach using priors based on an assumed density distribution of the Milky Way \citep[e.g.,][]{Bailer-Jones2018}, 4.485$\pm$0.89 kpc. It is important to note that our assumed Monte Carlo approach to compute the orbital elements are similars adopting both distance estimates, and therefore do not affect the results presented in this work.

For the Galactic model we employ the Galactic dynamic software GravPot16\footnote{\url{https://fernandez-trincado.github.io/GravPot16/}} (Fern\'andez-Trincado et al. 2018, in preparation), a semi-analytic, steady-state, three dimensional gravitational potential based on the mass density distributions of the Besan\c{c}on Galactic model  \citep{Robin_2003, Robin_2012, Robin_2014}, observationally and dynamically constrained. The model is constituted by seven thin disc components, two thick discs, an interstellar medium (ISM), a Hernquist stellar halo, a rotating bar component, and is surrounded by a spherical dark matter halo component that fits fairly well the structural and dynamical parameters of the Milky Way to the best we know them. A description of this model and its updated parameters appears in a score of papers \citep{Fernandez-Trincado_2016,Fernandez-Trincado_2017c,Fernandez-Trincado_2017b,Fernandez-Trincado_2017a,Tang_2017, Tang2018, Libralato_2018}.

The Galactic potential is scaled to the Sun's galactocentric distance, 8.3$\pm$0.23 kpc, and the local rotation velocity, 239$\pm$7 km s$^{1}$ \citep[e.g.,][]{Brunthaler_2011}. We assumed the Sun's orbital velocity vector [U$_{\odot}$,V$_{\odot}$,W$_{\odot}$]=[11.1$^{+0.69}_{-0.75}$, 12.24$^{+0.47}_{-0.47}$, 7.25$^{+0.37}_{-0.36}$] \citep{Schonrich_2010}. A long list of studies in the literature has presented different ranges for the bar pattern speeds. For our computations, the values $\Omega_{bar} = $ 35, 40, 45, and 50 km s$^{-1}$ kpc$^{-1}$ are employed. These values are consistent with the recent estimate of $\Omega_{bar}$ given by \cite{Portail_2017, Monari_2017a, Monari_2017b, Fernandez-Trincado_2017b}. We consider an angle of $\phi = 20^{\circ}$ for the present-day orientation of the major axis of the Galactic bar and the Sun-Galactic center line. The total mass of the bar taken in this work is 1.1$\times$10$^{10}$ M$_{\odot}$, which corresponds to the dynamical constraints towards the Milky Way bulge from massless particle simulations \cite[][]{Fernandez-Trincado_2017b} and is consistent with the recent estimate given by \citet{Portail_2017}. 

\begin{table}
\centering
\caption{ Orbital elements of IC 166 estimated with the newly-measured proper motions and parallax from \textit{Gaia} DR2 data combined with existing line-of-sight velocities from APOGEE. The average value of the orbital parameters of IC 166 was found for one mllion realizations adopting a Monte Carlo approach, with uncertainty ranges given by the 16th (subscript) and 84th (superscript) percentile values.}
\label{tab:blabla}
\begin{tabular}{ccccc}
\hline 
\hline
 ${\Omega}_{\rm bar}$        & $\langle$$r_{\rm peri}$$\rangle$  &  $\langle$$r_{\rm apo}$$\rangle$ & $\langle$$|z|_{\rm max}$$\rangle$ &  $\langle$$e$$\rangle$  \\ 
 (km s$^{-1}$ kpc$^{-1}$) & (kpc) & (kpc) & (kpc) &  \\ 
         \hline
  35  &  12.44$^{14.09}_{10.80}$ &  16.51$^{20.88}_{12.74}$ &  1.49$^{2.34}_{0.68}$ &  0.13$^{0.19}_{0.07}$ \\
  40  &  12.45$^{14.09}_{11.01}$ &  16.49$^{20.82}_{11.98}$ &  1.49$^{2.34}_{0.65}$ &  0.13$^{0.20}_{0.06}$ \\
  45  &  12.47$^{14.08}_{11.05}$ &  16.45$^{20.87}_{11.85}$ &  1.49$^{2.34}_{0.64}$ &  0.12$^{0.20}_{0.05}$ \\
  50  &  12.48$^{14.09}_{11.01}$ &  16.42$^{20.88}_{11.95}$ &  1.49$^{2.34}_{0.65}$ &  0.12$^{0.20}_{0.05}$ \\
   \hline
   \hline
\end{tabular}
\end{table}

The probable orbit of IC 166 is computed adopting a simple Monte Carlo procedure for different bar pattern speeds as mentioned above. For each of 10$^{3}$ simulations, we time-integrated backwards  the orbits for 2.5 Gyr under variations of the initial conditions (proper motions, radial velocity, heliocentric distance, Solar position, Solar motion and the velocity of the local standard of rest) according to their estimated errors, where the errors are assumed to follow a Gaussian distribution. The results of these computations are showed in Figure \ref{fig:sim_M1}. The same figures display the probability densities of the resulting orbits projected on the meridional and equatorial Galactic planes in the non-inertial reference frame where the bar is at rest. The yellow and red colors correspond to more probable regions of the space, which are crossed more frequently by the simulated orbits. The final point of each of these orbits has a very similar position to the current one of IC 166.

The median values of the orbital elements for the 10$^{3}$ realizations are listed in Table \ref{tab:blabla}. Uncertainties in the orbital integrations are estimated as the 16th (lower limit) and 84th (upper limit) percentile values. We defined the orbital eccentricity as: $$e = \frac{(r_{apo} - r_{peri})}{(r_{apo} + r_{peri} )},$$ where $r_{apo}$ is the apogalactic distance and $r_{peri}$ the perigalactic distance.  We find the orbit of IC 166 lies in the Galactic disc and it appears to be an unremarkable typical Galactic open cluster. 

\section{Conclusions}
\label{sec:sec7}

We have presented the first high resolution spectroscopic observations of the stellar cluster IC 166, which was recently surveyed in the \textit{H}-band of APOGEE. Based on their sky distribution, radial velocity, metallicity, CMD position and proper motions, we have identified 13 highest likelihood cluster members. We derived for the first time manual abundance determinations for up to 8 chemical species (Mg, Ca, Ti, Si, Al, K, Fe, Mn). High-resolution spectra are consistent with the cluster having a metallicity of [Fe/H]$=$ -0.08$\pm$0.05 dex. Isochrone fits indicate that the cluster is about 1.0$\pm$0.2 Gyr in age.

The results presented here show the cluster lies in the low-$\alpha$ sequence near the solar neighborhood, i.e., the cluster lies in the locus dominated by the low-$\alpha$ sequence of the canonical thin disc. We also found excellent agreement between our chemical abundances and general Galactic trends from large scale studies.

It is important to note that our manual analysis was able to reduce the dispersion found by APOGEE/ASPCAP pipeline for most of the chemical species studied in this work. The most notable improvement was for [Al/Fe] abundance ratios. 

Lastly, numerical integration of the possible orbits of IC 166 shows that the cluster appears to be an unremarkable standard Galactic open cluster with an orbit bound to the Galactic plane. The maximum and minimum Galactic distance achieved by the cluster as well as its orbital eccentricity suggest star formation at large Galactocentric radii. These results suggest that IC 166 could have formed nearer the solar neighborhood, fully compatible with the majority of known Galactic open clusters at similar metallicity. However, the derived orbital eccentricity ($\sim 0.13$) of the cluster is found be compatible with thin disc populations, but the maximum height above the plane, Zmax, larger than 1.5 kpc like IC 166 is too high for the thin disc and more compatible with the thick disc. It is important to note that, because the orbital excursions in our simulations are in the external part of the Galaxy (up to 16.5 kpc), it is in a region where the disc of the Milky Way is know to exhibit a significant flare \citep[e.g.,][]{Reyle2009} and warp \citep{Momany2006, Carraro_2007}. Such dynamical behaviour have been also observed in anti-center old open clusters, like Gaia 1 \citep[e.g.,][]{Koposov2017, Koch2018, Carraro2018}. 
	
We further note some important limitations of our orbital calculations: we ingore secular changes in the Milky Way potential over time. We also ignore the fact that the Milky Way disc exibit a prominent warp and flare in the direction of IC 166. The Milky Way potential that we used in the simulations is made-up of the seven time independent thin discs \citep{Robin_2003} with Einasto laws \citep{Einasto1979}.

\section*{Acknowledgements}

We would like to thank John Donor for helpful comments in the manuscript. We are grateful to the referee for a prompt and constructive report. J.G.F-T is supported by FONDECYT No. 3180210. J.S-U and D.G. gratefully acknowledge support from the Chilean BASAL Centro de Excelencia en Astrof\'isica y Tecnolog\'ias Afines (CATA) grant PFB-06/2007. B.T. acknowledges support from the one-hundred-talent project of Sun Yat-Sen University. SV gratefully acknowledges the support provided by Fondecyt reg n. 1170518. D.M. is supported by the BASAL Center for Astrophysics and Associated Technologies (CATA) through grant PFB-06, by the Ministry for the Economy, Development and Tourism,  Programa Iniciativa Cient\'ifica Milenio grant IC120009, awared to the Millennium Institute of Astrophysics (MAS), and by FONDECYT Regular grant No. 1170121. SzM has been supported by the Premium Postdoctoral Research Program of the Hungarian Academy of Sciences, and by the Hungarian NKFI Grants K-119517 of the Hungarian National Research, Development and Innovation Office. P.M.F. acknowledges support by an National Science Foundation AAG grants AST-1311835 \& AST-1715662.  V.V.S. and K.c. acknowledge support from NASA grant NNX17AB64G. OZ, FD, TM, and DAGH acknowledge support provided by the Spanish Ministry of Economy and Competitiveness (MINECO) under grant AYA-2014-58082-P. 

Funding for the \textit{GravPot16} software has been provided by the Centre national d'\'etudes spatiales (CNES) through grant 0101973 and UTINAM Institute of the 
Universit\'e de Franche-Comt\'e, supported by the R\'egion de Franche-Comt\'e and Institut des Sciences de l'Univers (INSU). Simulations have been executed on computers 
from the Utinam Institute of the Universit\'e de Franche-Comt\'e, supported by the R\'egion de Franche-Comt\'e and Institut des Sciences de l'Univers (INSU), and on 
the supercomputer facilities of the M\'esocentre de calcul de Franche-Comt\'e.

Funding for the Sloan Digital Sky Survey IV (SDSS-IV) has been provided by the Alfred P. Sloan Foundation, the US Department of Energy Office of Science, and the 
Participating Institutions. SDSS-IV acknowledges support and resources from the Center for High-Performance Computing at the University of Utah. The SDSS website is 
www.sdss.org.\\
SDSS-IV is managed by the Astrophysical Research Consortium for the Participating Institutions of the SDSS Collaboration including the Brazilian Participation Group, 
the Carnegie Institution for Science, Carnegie Mellon University, the Chilean Participation Group, the French Participation Group, Harvard-Smithsonian Center for 
Astrophysics, Instituto de Astrof\'isica de Canarias, The Johns Hopkins University, Kavli Institute for the Physics and Mathematics of the Universe (IPMU)/University 
of Tokyo, Lawrence Berkeley National Laboratory, Leibniz Institut f\"{u}r Astrophysik Potsdam (AIP), Max-Planck-Institut f\"{u}r Astronomie (MPIA Heidelberg), 
Max-Planck-Institut f\"{u}r Astrophysik (MPA Garching), Max-Planck-Institut f\"{u}r Extraterrestrische Physik (MPE), National Astronomical Observatory of China, 
New Mexico State University, New York University, University of Notre Dame, Observatorio Nacional/MCTI, The Ohio State University, Pennsylvania State University, 
Shanghai Astronomical Observatory, United Kingdom Participation Group, Universidad Nacional Aut\'onoma de M\'exico, University of Arizona, University of Colorado 
Boulder, University of Oxford, University of Portsmouth, University of Utah, University of Virginia, University of Washington, University of Wisconsin, Vanderbilt 
University, and Yale University.

\bibliographystyle{apj}{}
\bibliography{references}

\clearpage
\appendix
\section{Elemental abundances line-by-line}
\label{sec:apendice}
\setcounter{table}{0}
\renewcommand{\thetable}{A\arabic{table}}

\begin{table*}
    \centering    
    \caption{Atomic Lines used and Derived Abundances.}
    \label{tab:result_ferre}
\resizebox{\textwidth}{!}{\begin{tabular}{ccccccccccccccc}
\hline
\hline
Element & $\lambda_{air}$ & star1 & star2 & star3 & star4 & star5 & star6 & star7 & star8 & star9 & star10 & star11 & star12 & star13 \\ \hline
Fe& 15194.50  &  7.34  &  7.46  &  7.31  &  7.32  &  7.50  &  7.56  &  ...  &  7.43  &  7.35  &  ...  &  ...  &  7.39  &  7.40 \\
  & 15207.50  &  7.36  &  7.55  &  7.22  &  7.36  &  7.40  &  7.52  &  7.40  &  7.36  &  7.33  &  7.46  &  7.23  &  7.37  &  7.29 \\
  & 15490.30  &  7.45  &  7.40  &  7.49  &  7.53  &  7.54  &  7.47  &  7.40  &  ...  &  7.32  &  ...  &  7.43  &  7.50  &  ... \\
  & 15648.50  &  7.36  &  7.42  &  ...  &  ...  &  7.34  &  7.45  &  7.37  &  7.18  &  7.34  &  ...  &  ...  &  7.33  &  7.28 \\
  & 15964.90  &  7.28  &  7.58  &  7.41  &  7.37  &  7.40  &  7.46  &  7.34  &  ...  &  7.29  &  7.48  &  7.15  &  7.42  &  7.25 \\
  & 16040.70  &  7.29  &  ...  &  ...  &  7.54  &  7.42  &  7.28  &  7.49  &  7.32  &  7.36  &  7.47  &  7.31  &  7.29  &  7.37 \\
  & 16153.20  &  7.28  &  7.41  &  7.27  &  7.35  &  7.38  &  7.40  &  7.36  &  ...  &  7.21  &  7.48  &  7.26  &  7.32  &  7.25 \\
  & 16165.00  &  7.33  &  7.29  &  7.46  &  ...  &  7.33  &  7.26  &  7.32  &  ...  &  7.36  &  7.34  &  ...  &  7.36  &  ... \\ \hline
  &  $\langle$A(Fe)$\rangle$  &  7.34$\pm$0.06  &  7.44$\pm$0.10  &  7.36$\pm$0.11  &  7.41$\pm$ 0.10 &  7.41$\pm$0.07  &  7.42$\pm$0.11  &  7.38$\pm$0.05  &  7.32$\pm$0.10  &  7.32$\pm$0.05  &  7.45$\pm$0.06  &  7.28$\pm$0.10  &  7.37$\pm$0.06  &  7.31$\pm$0.06 \\ \hline
  & [Fe/H]   &  -0.11$\pm$0.06  &  -0.01$\pm$0.10  &  -0.09$\pm$0.11  &  -0.04$\pm$0.10  &  -0.04$\pm$0.07  &  -0.03$\pm$0.11  &  -0.07$\pm$0.05  &  -0.13$\pm$0.10  &  -0.13$\pm$0.05  &  0.00$\pm$0.06  &  -0.17$\pm$0.10  &  -0.08$\pm$0.06  &  -0.14$\pm$0.06 \\ \hline

Mg& 15740.70  &  7.27  &  7.27  &  7.31  &  7.30  &  7.33  &  7.28  &  7.37  &  7.20  &  7.23  &  7.38  &  7.19  &  7.36  &  7.19 \\
  & 15748.90  &  7.25  &  7.30  &  7.28  &  7.25  &  7.29  &  7.30  &  7.28  &  7.15  &  7.14  &  7.36  &  7.34  &  7.28  &  7.26 \\
  & 15765.80  &  7.26  &  7.23  &  7.26  &  7.23  &  7.29  &  7.34  &  7.23  &  7.15  &  7.17  &  7.31  &  ...  &  7.10  &  7.20 \\ \hline
  &  $\langle$A(Mg)$\rangle$  &  7.26$\pm$0.01  &  7.27$\pm$0.03  &  7.28$\pm$0.02  &  7.26$\pm$0.04  &  7.30$\pm$0.02  &  7.31$\pm$0.03  &  7.29$\pm$0.07  &  7.17$\pm$0.03  &  7.18$\pm$0.04  &  7.35$\pm$0.04  &  7.26$\pm$0.11  &  7.32$\pm$0.06  &  7.22$\pm$0.04   \\ \hline
  &  [Mg/Fe]  &  -0.16$\pm$0.01  &  -0.25$\pm$0.03  &  -0.16$\pm$0.02  &  -0.23$\pm$0.04  &  -0.19$\pm$0.02  &  -0.19$\pm$0.03  &  -0.17$\pm$0.07  &  -0.23$\pm$0.03  &  -0.22$\pm$0.04  &  -0.18$\pm$0.04  &  -0.10$\pm$0.11  &  -0.13$\pm$0.06  &  -0.17$\pm$0.04 \\ \hline

Ca& 16136.80  &  ...  &  6.07  &  6.04  &  6.17  &  6.14  &  6.09  &  ...  &  ...  &  ...  &  ...  &  ...  &  ...  &  ... \\
  & 16150.80  &  6.05  &  6.29  &  6.30  &  6.17  &  6.26  &  6.28  &  ...  &  ...  &  6.04  &  6.28  &  ...  &  6.18  &  6.11 \\
  & 16157.40  &  6.13  &  ...  &  ...  &  6.25  &  6.29  &  ...  &  ...  &  ...  &  6.30  &  6.18  &  ...  &  ...  &  ... \\
  & 16197.10  &  6.25  &  6.29  &  ...  &  6.34  &  6.31  &  ...  &  ...  &  ...  &  ...  &  ...  &  ...  &  6.36  &  ... \\ \hline
  &  $\langle$A(Ca)$\rangle$  &  6.14$\pm$0.10  &  6.22$\pm$0.13  &  6.17$\pm$0.18  &  6.23$\pm$0.08  &  6.25$\pm$0.08  &  6.18$\pm$0.13  &  ...  &  ...  &  6.17$\pm$0.18  &  6.23$\pm$0.07  &  ...  &  6.27$\pm$0.13  &  6.11 \\ \hline
  &  [Ca/Fe]  &  -0.06$\pm$0.10  &  -0.08$\pm$0.13  &  -0.05$\pm$0.18  &  -0.04$\pm$0.08  &  -0.02$\pm$0.08  &  -0.10$\pm$0.13  &  ...  &  ...  &  -0.01$\pm$0.18  &  -0.08$\pm$0.07  &  ...  &  0.04$\pm$0.13  &  -0.06 \\ \hline

K& 15163.10  &  4.85  &  5.01  &  5.08  &  ...  &  5.12  &  ...  &  5.02  &  ...  &  ...  &  ...  &  ...  &  ...  &  ... \\
  & 15168.40  &  4.99  &  5.08  &  4.97  &  ...  &  5.15  &  4.93  &  ...  &  4.98  &  ...  &  ...  &  ...  &  5.14  &  ... \\ \hline
  &  $\langle$A(K)$\rangle$  &  4.92$\pm$0.10  &  5.04$\pm$0.05  &  5.02$\pm$0.08  &  ...  &  5.13$\pm$0.02  &  4.93  &  5.02  &  4.98  &  ...  &  ...  &  ...  &  5.14  &  ... \\ \hline
  &  [K/Fe]  &  -0.05$\pm$0.10  &  -0.03$\pm$0.05  &  0.03$\pm$0.08  &  ...  &  0.04$\pm$0.02  &  -0.12  &  0.01  &  0.03  &  ...  &  ...  &  ...  &  0.14  &  ... \\ \hline

Si& 15376.80  &  7.32  &  7.46  &  ...  &  ...  &  7.45  &  7.59  &  ...  &  7.40  &  ...  &  ...  &  ...  &  7.46  &  ... \\
  & 15557.80  &  7.42  &  7.63  &  7.39  &  7.58  &  7.54  &  7.59  &  7.39  &  7.42  &  ...  &  7.55  &  ...  &  7.27  &  7.36 \\
  & 15884.50  &  7.29  &  7.45  &  ...  &  7.32  &  7.44  &  7.40  &  7.33  &  7.29  &  7.33  &  7.38  &  7.21  &  7.31  &  7.19 \\
  & 15960.10  &  7.64  &  7.73  &  7.55  &  7.54  &  7.73  &  7.59  &  7.59  &  ...  &  7.55  &  7.59  &  7.38  &  7.88  &  7.55 \\
  & 16060.00  &  ...  &  ...  &  ...  &  7.79  &  ...  &  ...  &  ...  &  7.45  &  ...  &  ...  &  7.66  &  7.53  &  7.52 \\
  & 16094.80  &  ...  &  7.58  &  7.53  &  ...  &  7.52  &  7.43  &  7.43  &  ...  &  7.50  &  7.54  &  ...  &  7.61  &  7.45 \\
  & 16215.70  &  ...  &  7.71  &  ...  &  7.55  &  7.64  &  7.73  &  ...  &  ...  &  7.49  &  7.74  &  7.55  &  ...  &  ... \\
  & 16241.80  &  ...  &  7.71  &  ...  &  7.64  &  ...  &  ...  &  7.56  &  7.49  &  7.56  &  ...  &  7.51  &  7.36  &  ... \\
  & 16680.80  &  7.51  &  7.38  &  7.45  &  ...  &  7.61  &  7.45  &  7.51  &  7.41  &  7.39  &  7.57  &  7.42  &  7.42  &  7.51 \\
  & 16828.20  &  ...  &  ...  &  ...  &  ...  &  ...  &  ...  &  ...  &  7.55  &  7.37  &  ...  &  ...  &  ...  &  ... \\ \hline
  &  $\langle$A(Si)$\rangle$  &  7.44$\pm$0.14  &  7.58$\pm$0.14  &  7.48$\pm$0.07  &  7.57$\pm$0.15  &  7.56$\pm$0.10  &  7.54$\pm$0.12  &  7.48$\pm$0.10  &  7.43$\pm$0.08  &  7.45$\pm$0.09  &  7.56$\pm$0.11  &  7.41$\pm$0.13  &  7.48$\pm$0.20  &  7.43$\pm$0.13  \\ \hline
  &  [Si/Fe]  &  0.04$\pm$0.14  &  0.08$\pm$0.14  &  0.06$\pm$0.07  &  0.10$\pm$0.15  &  0.09$\pm$0.10  &  0.06$\pm$0.12  &  0.04$\pm$0.10  &  0.05$\pm$0.08  &  0.07$\pm$0.09  &  0.05$\pm$0.11  &  0.07$\pm$0.13  &  0.05$\pm$0.20  &  0.06$\pm$0.13 \\ \hline

Ti  & 15715.60  &  4.71  &  4.78  &  ...  &  ...  &  ...  &  ...  &  ...  &  ...  &  ...  &  4.84  &  ...  &  ...  &  4.78 \\ \hline
  &  $\langle$A(Ti)$\rangle$  &  4.71  &  4.78  &  ...  &  ...  &  ...  &  ...  &  ...  &  ...  &  ...  &  4.84  &  ...  &  ...  &  4.78 \\ \hline
  &  [Ti/Fe]  &  -0.08  &  -0.11  &  ...  &  ...  &  ...  &  ...  &  ...  &  ...  &  ...  &  -0.06  &  ...  &  ...  &  0.02 \\ \hline

Mn& 15159.20  &  5.26 &  5.28  &  ...  &  5.28  &  ...  &  5.34  &  ...  &  5.26  &  ...  &  5.34  &  ...  &  5.31  &  ... \\
  & 15217.70  &  5.23 &  5.38  &  5.37  &  5.25  &  5.37  &  5.32  &  5.31  &  5.23  &  5.19  &  5.34  &  5.27  &  ...  &  ... \\
  & 15262.40  &  ...  &  5.29  &  5.20  &  5.37  &  ...  &  ...  &  5.25  &  5.29  &  5.32  &  5.35  &  5.30  &  5.33  &  ... \\ \hline
  &  $\langle$A(Mn)$\rangle$  &  5.24$\pm$0.02  &  5.32$\pm$0.05  &  5.28$\pm$0.12  &  5.30$\pm$0.06  &  5.37  &  5.33$\pm$0.01  &  5.28$\pm$0.04  &  5.26$\pm$0.03  &  5.25$\pm$0.09  &  5.34$\pm$0.01  &  5.28$\pm$0.02  &  5.32$\pm$0.01  &  ...  \\ \hline
  &  [Mn/Fe]  &  -0.04$\pm$0.02  &  -0.06$\pm$0.05  &  -0.02$\pm$0.12  &  -0.05$\pm$0.06  &  0.02  &  -0.03$\pm$0.01  &  -0.04$\pm$0.04  &  0.00$\pm$0.03  &  -0.01$\pm$0.09  &  -0.05$\pm$0.01  &  0.06$\pm$0.02  &  0.01$\pm$0.01  &  ... \\ \hline

Al& 16719.00  &  6.51 &  6.45  &  6.47  &  6.47  &  ...  &  ...  &  6.43  &  6.38  &  6.45  &  6.51  &  ...  &  6.39  &  ... \\
  & 16750.60  &  6.37 &  6.41  &  6.45  &  6.37  &  6.50  &  6.48  &  6.32  &  6.18  &  6.36  &  6.39  &  6.33  &  6.37  &  ... \\ \hline
  &  $\langle$A(Al)$\rangle$  &  6.44$\pm$0.10  &  6.43$\pm$0.03  &  6.46$\pm$0.01  &  6.42$\pm$0.07  &  6.50  &  6.48  &  6.37$\pm$0.08  &  6.28$\pm$0.14  &  6.40$\pm$0.06  &  6.45$\pm$0.08  &  6.33  &  6.38$\pm$0.01  &  ... \\ \hline
  &  [Al/Fe]  &  0.18$\pm$0.10  &  0.07$\pm$0.03  &  0.18$\pm$0.01  &  0.09$\pm$0.07  &  0.17  &  0.14  &  0.07$\pm$0.08  &  0.04$\pm$0.14  &  0.16$\pm$0.06  &  0.07$\pm$0.08  &  0.13  &  0.08$\pm$0.01  &  ... \\
\hline
\hline
\end{tabular}}
    \end{table*}

\newpage

\section{Orbit of IC 166 with Monte Carlo calculations}

\begin{figure*}[h!]
	\centering
\includegraphics[width=0.80\textwidth]{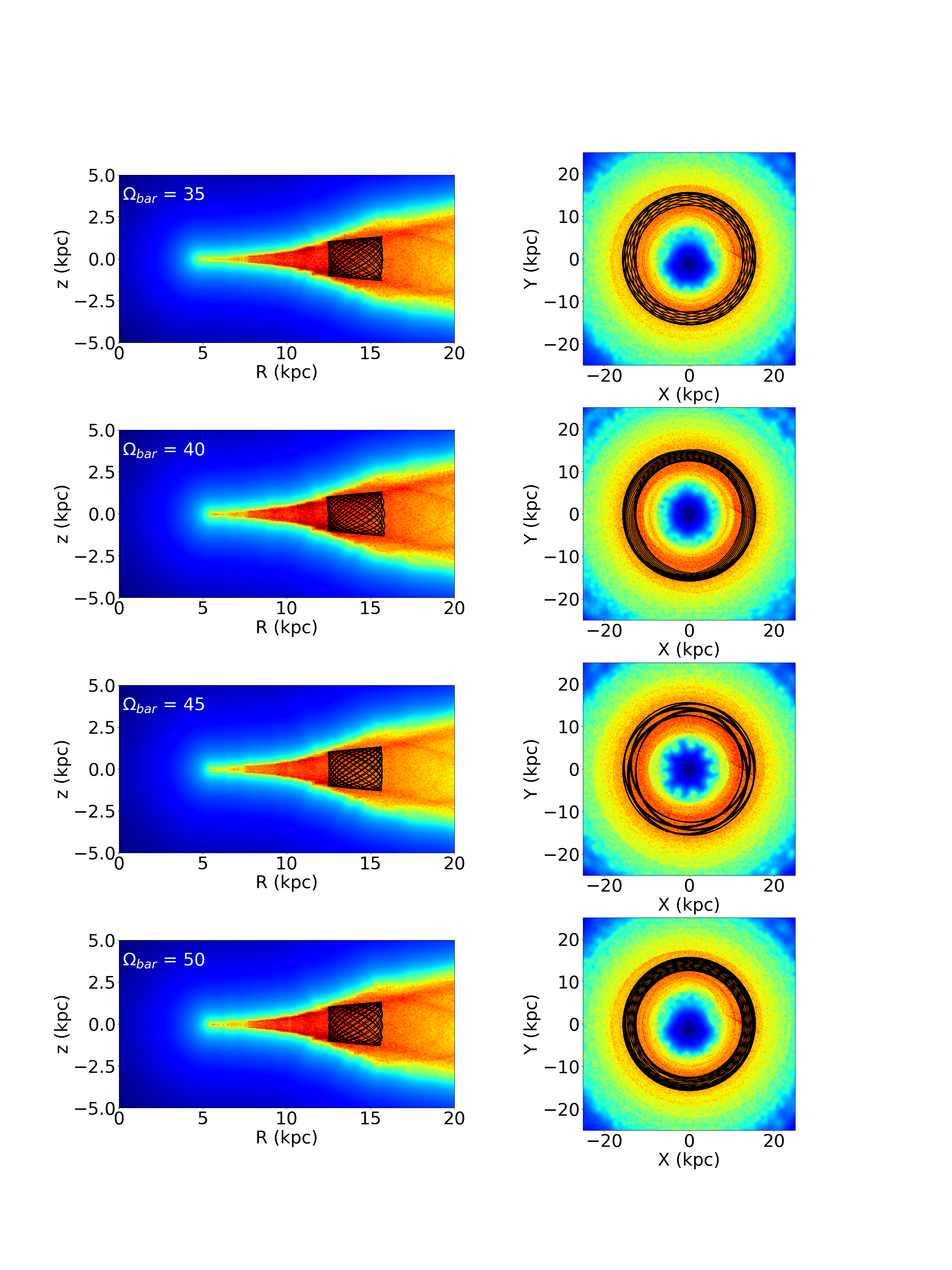}
    \caption{Probability density maps color-coded at the bottom for the meridional orbits in the $R,z$ plane (column 1) and face-on (column 2) of a thousand random realizations of IC 166 time-integrated backwards for 2.5 Gyr adopting the newly-measured proper motions and parallax from \textit{Gaia} DR2 \citep{Gaia2018, Lindegren2018} . Red and yellow colors correspond to larger probabilities. The tile size of the HealPix map is 0.10 kpc$^{2}$. The black line shows the orbit using the best values found for the cluster (see text).}
    \label{fig:sim_M1}
\end{figure*}

Figure \ref{fig:sim_M1} shows the Monte Carlo simulations for the bound orbit of IC 166, we make these Monte Carlo simulations to estimate the uncertainties in the orbital elements (see text).

\end{document}